\documentclass[journal=jctcce,manuscript=article,layout=twocolumn]{achemso}

\usepackage[utf8]{inputenc}
\usepackage[T1]{fontenc}
\usepackage[final]{microtype}
\usepackage[american]{babel}
\usepackage{graphicx}
\usepackage{booktabs}
\usepackage{enumitem}

\usepackage[version=4]{mhchem}
\usepackage{siunitx}
\usepackage{amsmath}

\DeclareSIUnit\angstrom{\text{Å}}
\DeclareSIUnit\kcal{\text{kcal}}
\DeclareSIUnit\bar{\text{bar}}
\author{Rinto Thomas}
\affiliation[UMR]{Fachbereich Chemie, Philipps-Universität Marburg, 35032 Marburg, Germany}
\altaffiliation{R.T. and P.R.P. contributed equally to this work.}
\author{Praveen Ranganath Prabhakar}
\affiliation[UCI]{Department of Chemistry, University of California, Irvine, Irvine, California 92697, United States}
\altaffiliation{R.T. and P.R.P. contributed equally to this work.}
\author{Michael von Domaros}
\affiliation[UMR]{Fachbereich Chemie, Philipps-Universität Marburg, 35032 Marburg, Germany}
\email{mvondomaros@uni-marburg.de}

\title{A Residence-Time Approach for Determining Position-Dependent Diffusivities from Biased Molecular Simulations}

\begin{document}

\begin{tocentry}
    \includegraphics[width=3.25in]{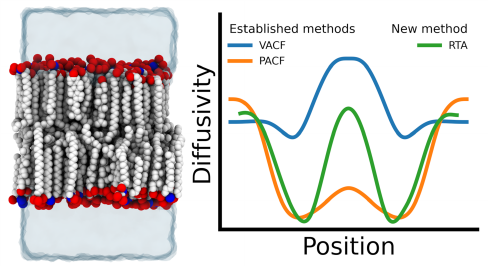}
\end{tocentry}

\begin{abstract}
Position-dependent diffusivities are central parameters in reduced stochastic descriptions of molecular transport in heterogeneous environments, but their reliable estimation from molecular dynamics simulations remains challenging. We present a residence-time approach (RTA) that extracts local diffusivities from first-exit statistics measured in biased simulations after compensation of the mean free-energy gradient. We apply the method to oxygen diffusion across a hexadecane/water slab, water permeation across a POPC lipid bilayer, and transport of water and volatile organic compounds through a model skin-barrier membrane. In the slab system, RTA diffusivities agree with independently determined bulk reference values. In the membrane systems, propagator predictions based on RTA-derived diffusivities reproduce unbiased molecular dynamics propagators over substantial lag-time ranges, while also revealing that, in some cases, no single lag-time-independent diffusivity profile captures the dynamics across all timescales. These results support residence-time statistics as a practical route for determining effective position-dependent diffusivities from biased molecular simulations.
\end{abstract}

\section{Introduction}

Diffusivities provide a compact and physically interpretable description of complex many-body dynamics, whether
generated by molecular dynamics (MD) simulations or observed in natural systems, in terms of effective transport
coefficients.\cite{Einstein1905,Kubo1957} By averaging over microscopic degrees of freedom, a diffusivity captures
how interactions and solvent-induced fluctuations determine transport on mesoscopic scales. In homogeneous
environments, this coarse-grained description reduces to a single scalar diffusivity. In heterogeneous systems such
as membranes, nanopores, and porous materials, however, transport is more naturally described by a
position-dependent diffusivity $D(z)$ along a chosen transport coordinate $z$.\cite{Hummer2005}

In a continuum description, position-dependent diffusivities arise naturally within stochastic transport
models.\cite{RiskenHaken1996} Within this framework, the probability density $p(z,t)$ evolves according to the
Smoluchowski equation,
\begin{equation}
\partial_t p(z,t)
=
\partial_z\!\left[
D(z)\, e^{-\beta F(z)}
\,\partial_z\!\left(e^{\beta F(z)} p(z,t)\right)
\right],
\label{eq:smoluchowski}
\end{equation}
where $p(z,t)$ is the probability density and $F(z)$ is the potential of mean force (PMF) along $z$, defined
relative to the bulk phase such that $F(z_{\mathrm{bulk}})=0$. In this effective one-dimensional description,
$F(z)$ and $D(z)$ encode the thermodynamic and kinetic consequences of projecting high-dimensional dynamics onto a
reduced coordinate. Eq.~\eqref{eq:smoluchowski} represents a diffusive, memoryless (Markovian) description of the
projected dynamics along $z$.

A particularly important application is membrane permeation, where the steady-state solution of
Eq.~\eqref{eq:smoluchowski} yields the inhomogeneous solubility--diffusion (ISD)
model,\cite{DiamondKatz1974,VenableKramer2019}
\begin{equation}
P^{-1}
=
\int_{z_1}^{z_2}\!
\frac{e^{\beta F(z)}}{D(z)}\,dz,
\label{eq:isd}
\end{equation}
which relates the permeability $P$ to the PMF $F(z)$ and diffusivity $D(z)$ between boundaries $z_1$ and $z_2$
that delimit the membrane region. This relation provides a direct quantitative bridge between microscopic
simulations and experimentally measurable permeabilities, provided that both $F(z)$ and $D(z)$ are determined with sufficient accuracy.

Accurate determination of $D(z)$ is therefore essential for connecting molecular-scale dynamics to macroscopic transport properties. Extracting reliable position-dependent diffusivities from MD trajectories, however, remains challenging, and a wide range of approaches has been developed.\cite{LeeComer2016,AwoonorWilliamsRowley2016,Shinoda2016,VenableKramer2019} Broadly, these methods include equilibrium fluctuation-based estimators based on restrained trajectories and time-correlation functions,\cite{WoolfRoux1994,Hummer2005,GaalswykAwoonorWilliams2016,StraubBorkovec1987} likelihood-based inference of Smoluchowski dynamics from short-time propagators,\cite{Hummer2005,GhyselsVenable2017,KramerGhysels2020,ComerChipot2013} local-statistics approaches that estimate diffusivities from drift-free trajectory segments or local mean-squared displacements within finite observation domains,\cite{Nagai2020PositiondependentDiffusion,Kikkawa2026LocalDiffusion} discrete-state kinetic reconstructions such as Markov state models and milestoning,\cite{RostaHummer2015,StelzlKells2017} and short-lag moment or Kramers--Moyal estimators.\cite{SicardKoskin2021} Among these, fluctuation-based estimators obtained from harmonically restrained simulations are by far the most widely used in practice.

Many of these estimators are sensitive to analysis choices and model assumptions. Recent studies suggest that external confinement can modify the effective friction and memory kernel experienced by a solute, implying that diffusivities extracted from harmonically restrained dynamics need not coincide with those obtained from freely diffusing trajectories.\cite{Daldrop2017ExternalPotential,Yamaguchi2021DecouplingSolvent} Consequently, discrepancies between different estimators may reflect differences in the effective dynamics induced by the underlying sampling protocol rather than solely methodological shortcomings. Correlation-function approaches furthermore require numerical integration or extrapolation of noisy time-correlation data and may depend strongly on restraint strength and on the chosen integration or extrapolation scheme.\cite{WoolfRoux1994,Hummer2005,GaalswykAwoonorWilliams2016} Inference-based schemes can exhibit lag-time dependence and may be affected by discretization, regularization, or parametrization bias.\cite{Hummer2005,KramerGhysels2020} Local-statistics approaches require the definition of a finite observation domain and appropriate treatment of finite-window effects,\cite{Nagai2020PositiondependentDiffusion,Kikkawa2026LocalDiffusion} whereas discrete-state approaches require suitable state decompositions and adequate transition statistics, and short-lag moment estimators rely on a well-resolved regime in which the projected dynamics are approximately Markovian.\cite{RostaHummer2015,StelzlKells2017,SicardKoskin2021}

More fundamentally, the projection of high-dimensional molecular motion onto a single reduced coordinate can
introduce residual memory effects and incomplete separation of time scales, especially in heterogeneous
environments such as membranes.\cite{ChipotComer2016,GhyselsRoet2021,VervustZhang2023} As a result, diffusivity
estimates may depend on lag time or on the operational definition of the reduced dynamics, reflecting deviations
from an idealized memoryless Smoluchowski description. In many applications, however, the primary goal is not to resolve non-Markovian effects explicitly, but to construct an effective diffusive description that reproduces the transport observables of interest. Such a description may remain accurate for these observables even if the Markovian approximation breaks down in parts of the system.

Here, we introduce a residence-time approach (RTA) for estimating position-dependent diffusivities from biased MD simulations. The central idea is to determine $D(z)$ from mean first-exit times out of predefined spatial intervals along a transport collective variable (CV) $z$. Similar to previous approaches based on Bayesian inference and local displacement statistics,\cite{ComerChipot2013,Nagai2020PositiondependentDiffusion,Kikkawa2026LocalDiffusion} the present method is formulated for trajectory segments in which the effective drift along the CV is negligible. In principle, such a drift-free regime can be realized using any enhanced-sampling method that provides an estimate of the underlying free-energy profile or mean force, allowing a compensating bias to flatten the effective free-energy landscape. In the present work, we employ adaptive biasing force (ABF) simulations, for which this condition arises naturally as the estimated mean force converges, providing a convenient transition from free-energy to diffusivity estimation.

Our estimator is designed to avoid several practical limitations of established fluctuation-based approaches. It does not require harmonic confinement or dedicated restrained simulations, and it avoids numerical integration or extrapolation of noisy time-correlation functions.\cite{WoolfRoux1994,Hummer2005,GaalswykAwoonorWilliams2016} Instead, it determines local diffusivities directly from residence-time statistics in an approximately drift-free landscape, making the method straightforward to implement and naturally compatible with blocking analysis for uncertainty estimation.

We assess the method in two complementary ways. First, where bulk reference diffusivities are accessible, we compare the RTA estimates directly against independently determined bulk values. Second, and more generally, we perform propagator-level validation in heterogeneous environments: diffusivity profiles inferred from the RTA are combined with the PMF to compute model propagators, i.e., time-dependent conditional probability distributions, which are then compared with the same quantity obtained from unbiased MD simulations. Comparison between model and simulation propagators over a range of lag times provides a direct test of how well the inferred Smoluchowski model captures the projected dynamics.

The remainder of this article is organized as follows. We first develop the theoretical basis of the RTA and its connection to the drift-free limit of biased dynamics, then describe its implementation for ABF trajectories and apply it to systems ranging from simple liquid--liquid diffusion to increasingly heterogeneous membrane environments.

\section{Theory}

\subsection{Drift-free limit of biased dynamics}
\label{sec:drift_free}

As introduced above, the projected dynamics along a transport coordinate $z$ are modeled here by the
Smoluchowski equation, Eq.~\eqref{eq:smoluchowski}, with position-dependent diffusivity $D(z)$ and PMF $F(z)$.
The residence-time approach developed below applies to trajectory segments in which the effective drift along
$z$ is negligible. This condition can be realized by applying a bias potential $V_{\mathrm b}(z)$ that compensates
the underlying free-energy gradient, so that the effective free-energy profile
\begin{equation}
F_{\mathrm{eff}}(z)=F(z)+V_{\mathrm b}(z)
\label{eq:feff}
\end{equation}
is approximately constant over the region analyzed.

In the present work, this regime is realized using adaptive biasing force (ABF)
sampling.\cite{DarvePohorille2001,ComerGumbart2015} ABF estimates the mean force along a CV
during the simulation and applies a history-dependent biasing force that progressively compensates this mean force.
In one dimension, the biasing force approaches
\begin{equation}
-\frac{\partial V_{\mathrm b}(z,t)}{\partial z}
\approx
\frac{dF(z)}{dz},
\label{eq:abf_bias_force}
\end{equation}
so that, as the estimate converges, $F_{\mathrm{eff}}(z)$ becomes approximately flat.
The residual effective drift then becomes small on the spatial scale over which the bias has converged.

In this drift-free limit, the Smoluchowski equation reduces locally to
\begin{equation}
\partial_t p(z,t)
=
\partial_z\!\left[D(z)\,\partial_z p(z,t)\right],
\label{eq:rta_flat_diffusion}
\end{equation}
which describes diffusion in the absence of an effective free-energy gradient. This is the central condition
underlying the residence-time approach. In practice, the analysis must therefore be restricted to regions and
trajectory intervals for which the ABF bias has converged sufficiently that residual drift is negligible on the
scale of the residence-time intervals.

\subsection{Residence-time approach}
\label{sec:rta}

\subsubsection{First-exit times in an interval}

Consider an interval $\Omega=[a,b)$ along $z$ with width $L=b-a$. For the derivation, we assume that the
diffusivity is constant within the interval, $D(z)=D$ for $z\in\Omega$. The first-exit time $T$ is the random time
at which a trajectory leaves $\Omega$ for the first time,
\begin{equation}
\label{eq:rta_first_exit}
T = \inf\{ t \ge 0 : z(t) \notin \Omega \}.
\end{equation}

For an initial condition $z(0)=z_0\in(a,b)$, the mean first-exit time (MFET)
\begin{equation}
\label{eq:rta_mfet_def}
\tau(z_0)\equiv \langle T \mid z(0)=z_0\rangle
\end{equation}
is the average time required to leave $\Omega$ when starting at $z_0$.

For drift-free Markovian diffusion with constant diffusivity in the interval, $\tau(z_0)$ satisfies a standard
backward equation.\cite{Redner2001,Gardiner2009} This result can be obtained by a one-step argument. Writing
\begin{equation}
\label{eq:rta_onestep}
\tau(z_0)=\delta t+\left\langle \tau\!\left(z_0+\delta z\right)\right\rangle,
\end{equation}
with $\delta z=z(\delta t)-z(0)$, and using $\langle \delta z\rangle=0$ and
$\langle (\delta z)^2\rangle=2D\,\delta t$ for drift-free diffusion, a second-order expansion in $\delta z$
yields
\begin{equation}
\label{eq:rta_mfet_bvp}
D\,\tau''(z_0) = -1,
\quad
\tau(a)=0,
\quad
\tau(b)=0.
\end{equation}
The absorbing boundary conditions express that the first-exit time vanishes when the trajectory starts at either
boundary. Solving Eq.~\eqref{eq:rta_mfet_bvp} gives
\begin{equation}
\label{eq:rta_mfet_solution}
\tau(z_0)=\frac{(z_0-a)(b-z_0)}{2D}.
\end{equation}

\subsubsection{Residence-time identity}

We define the residence time $\tau_r$ as the MFET averaged over initial positions $z_0$ within $\Omega$,
\begin{equation}
\label{eq:rta_residence_def}
\tau_r
=
\langle T \rangle_{\Omega}
=
\int_a^b \rho_{\Omega}(z_0)\,\tau(z_0)\,dz_0,
\end{equation}
where $\rho_{\Omega}(z_0)$ denotes the equilibrium distribution of positions conditioned on the particle being inside the interval. As in Eq.~\eqref{eq:rta_mfet_def}, $z_0$ denotes the initial position for a first-exit problem, not necessarily the position at which a trajectory enters the interval. In the trajectory analysis, each sampled frame for which the particle is inside $\Omega$ defines such a possible initial condition, and the corresponding residence time is the remaining first-exit time from that position. Re-entry into the interval is therefore accounted for through later sampled frames, rather than by extending a previous residence event.

For drift-free diffusion in a flat effective PMF, the conditional equilibrium distribution is uniform within the interval, so that $\rho_{\Omega}(z_0)=1/L$. Inserting Eq.~\eqref{eq:rta_mfet_solution} into Eq.~\eqref{eq:rta_residence_def} then gives
\begin{equation}
\label{eq:rta_identity}
\tau_r=\frac{L^2}{12D}.
\end{equation}
This relation is exact for one-dimensional drift-free diffusion with constant diffusivity inside the interval.

An equivalent derivation follows from solving the diffusion equation on $\Omega$ with absorbing boundaries and then integrating the corresponding survival probability over time to obtain the mean residence time.\cite{MercierFrancoCastier2016,HollringBaer2023}

\subsubsection{Local estimator for position-dependent diffusivities}

To estimate a position-dependent diffusivity, we discretize the transport coordinate into intervals
$\Omega_i=[z_i,z_{i+1})$ of width $L_i=z_{i+1}-z_i$. Within each interval, the diffusivity is approximated as
locally constant, $D(z)\approx D_i$ for $z\in\Omega_i$, such that spatial variations contribute only at higher
order in $L_i$.

Let $\tau_{r,i}$ denote the theoretical mean residence time in interval $\Omega_i$. By applying the
residence-time identity, Eq.~\eqref{eq:rta_identity}, locally to each interval, we obtain the estimator
\begin{equation}
\label{eq:rta_local_estimator}
\widehat{D}_i
=
\frac{L_i^2}{12\,\widehat{\tau}_{r,i}},
\end{equation}
where $\widehat{\tau}_{r,i}$ is the corresponding interval-averaged residence time estimated from the trajectory.

\begin{figure*}[t]
    \centering
    \includegraphics[width=\textwidth]{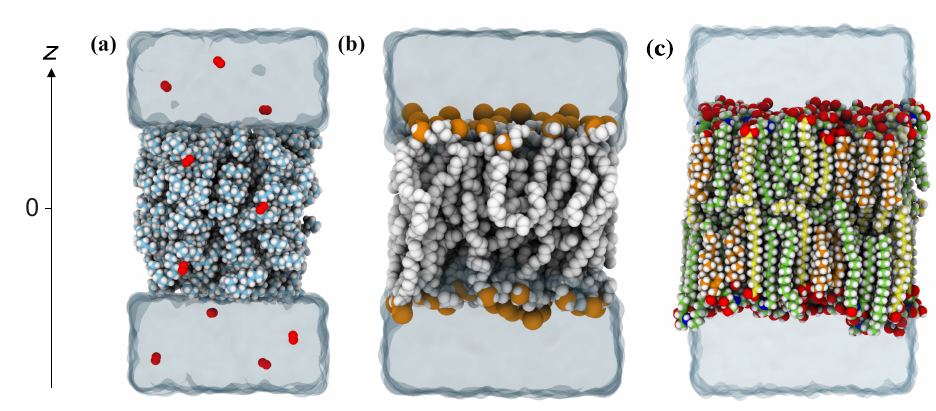}
    \caption{Representative snapshots of the model systems investigated in this work. (a) Hexadecane/water interface, with the diffusing oxygen atoms highlighted as red spheres. (b) POPC lipid bilayer. (c) Stratum corneum (SC) lipid matrix composed of ceramide NS (green), cholesterol (orange), and lignoceric acid (yellow). The transport coordinate $z$ and the reference position $z=0$, common to all three systems, are indicated on the left.}
    \label{fig:model_systems}
\end{figure*}

In practice, the residence-time identity is applied to biased MD trajectories only after the ABF bias has
converged sufficiently that the effective drift along the CV is negligible over the interval considered. The
procedure used to estimate $\widehat{\tau}_{r,i}$ from sampled trajectories is described in the Computational
Details section.

The resulting estimator rests on three central approximations. First, the
diffusivity is treated as locally constant within each interval. Second, the
effective drift is assumed to be negligible within the analyzed region, so that
the reduced dynamics are well approximated by
Eq.~\eqref{eq:rta_flat_diffusion}. Third, the projected dynamics are assumed to
be sufficiently close to Markovian on the spatial and temporal scales probed by
the residence-time analysis. Accordingly, the quantity
$\widehat{D}_i$ should be interpreted as an effective first-passage diffusivity
associated with the chosen residence interval rather than as a unique local
transport coefficient. The quality of these approximations is assessed below
through comparison to independent bulk diffusivities and, more rigorously,
through propagator-level validation in heterogeneous membrane systems.
Additional local MSD and survival-probability analyses are provided in
Section~S5.1, where this interpretation is further examined.

\section{Computational Details}

\subsection{Model Systems}

We evaluated the residence-time approach (RTA) for three systems of increasing complexity (Figure~\ref{fig:model_systems}): oxygen diffusion across a hexadecane/water slab, water permeation across a POPC lipid bilayer, and permeation of water, acetone, and 6-methyl-5-hepten-2-one (6-MHO) through a multicomponent lipid membrane representing the stratum corneum (SC) skin barrier. System compositions and simulation box dimensions are summarized in Table~\ref{tab:systems}.

\begin{table*}
\caption{Compositions of the simulated systems and corresponding simulation box dimensions.}
\label{tab:systems}
\centering
\begin{tabular}{lllll}
\toprule
System & Composition & $L_x$ (\si{\angstrom}) & $L_y$ (\si{\angstrom}) & $L_z$ (\si{\angstrom}) \\
\midrule
Hexadecane/water & 252 hexadecane, 4318 water, 10 \ce{O2} & 50.00 & 50.00 & 105.00 \\
POPC/water       & 72 POPC, 2967 water                    & 48.43 & 48.43 & 76.34  \\
SC/water         & 200 SC lipids, 4800 water             & 57.14 & 57.14 & 95.14  \\
SC/acetone       & 200 SC lipids, 4800 water, 1 acetone  & 57.14 & 57.14 & 95.35  \\
SC/6-MHO         & 200 SC lipids, 4800 water, 1 6-MHO    & 57.14 & 57.14 & 95.14  \\
\bottomrule
\end{tabular}

\vspace{0.5ex}
\raggedright
\footnotesize
SC lipids comprise an approximately equimolar mixture of cholesterol, lignoceric acid, and ceramide NS (24:0).\cite{ThomasPrabhakar2025}
For the POPC/water and SC/water systems, one water molecule was designated as the permeating solute for analysis.
\end{table*}

The hexadecane/water system serves as a simple liquid--liquid reference with low free-energy barriers and
well-defined bulk regions, enabling direct comparison with diffusivities obtained from the Einstein relation.
Our primary slab system is approximately twice as large as that of Ghysels \textit{et al.}\cite{GhyselsVenable2017},
yielding more extended bulk regions; a smaller comparison system more closely matching their setup is described in
Section~S1.4.

Water permeation across a POPC bilayer is a widely used benchmark for transport in fluid-phase lipid membranes and
has been investigated extensively in molecular dynamics (MD)
simulations.\cite{MathaiTristramNagle2007,NagleMathai2007,OrsiEssex2010,ComerSchulten2014,
AwoonorWilliamsRowley2016,VenableKramer2019}

The SC membrane represents a more complex and application-relevant system. Here we consider permeation of water,
acetone, and 6-MHO through a multicomponent SC lipid matrix, building on our previous work on the transport of
skin-oil oxidation products through the SC barrier.\cite{ThomasPrabhakar2025,ThomasPrabhakar2026}

All simulations employed the CHARMM36 lipid force field together with the CHARMM General Force Field (CGenFF) for
small molecules and the TIP3P water model.\cite{KlaudaVenable2010,VanommeslaegheHatcher2010,
JorgensenChandrasekhar1983} The oxygen model was adopted from Ref.~\cite{GhyselsVenable2017}.

Initial coordinates for the hexadecane/water system were generated using Packmol,\cite{MartinezAndrade2009}
whereas initial configurations for the POPC bilayer and SC membrane systems were taken from our previous
studies.\cite{ThomasPrabhakar2025,ThomasPrabhakar2026}

\subsection{Simulation Protocol}

All simulations were performed using NAMD~3.0.1.\cite{PhillipsBraun2005,PhillipsHardy2020} A multiple time-step
integration scheme based on the r-RESPA algorithm\cite{HumphreysFriesner1994} was employed with a base time step of
\qty{2}{\femto\second}. Short-range nonbonded interactions were evaluated every \qty{2}{\femto\second}, while
long-range electrostatic interactions were evaluated every \qty{4}{\femto\second}.

Water molecules were maintained as rigid bodies using the SETTLE algorithm,\cite{MiyamotoKollman1992} and all other bonds involving hydrogen atoms were constrained using the SHAKE
algorithm.\cite{RyckaertCiccotti1977}

Periodic boundary conditions were applied in all three spatial directions. Short-range nonbonded interactions were
switched between \qty{10}{\angstrom} and \qty{12}{\angstrom}. Long-range electrostatic interactions were computed
using the smooth particle mesh Ewald (SPME) method\cite{EssmannPerera1995} with a real-space cutoff of
\qty{12}{\angstrom}, a grid spacing of \qty{1}{\angstrom}, and sixth-order spline interpolation.

All simulations were carried out at \qty{310}{\kelvin}, except for the SC membrane system, which was simulated at
the physiological skin temperature of \qty{305.15}{\kelvin}. Temperature was controlled using the stochastic
velocity-rescaling thermostat of Bussi \emph{et al.}\cite{BussiDonadio2007} with a time constant of
\qty{1}{\pico\second}.

The POPC and SC membrane systems were initialized from equilibrated configurations obtained in previous studies,
where the preparation and equilibration procedures are described in
detail.\cite{ThomasPrabhakar2025,ThomasPrabhakar2026}

For the hexadecane/water system, energy minimization was performed for 100000 steps, followed by \qty{0.5}{\nano\second} of NVT equilibration. This was followed by \qty{40}{\nano\second} of equilibration at \qty{1}{\bar} with the lateral box dimensions fixed (constant area), using a Langevin piston barostat\cite{MartynaTobias1994,FellerZhang1995} with a relaxation time of \qty{50}{\femto\second} and an oscillation period of \qty{100}{\femto\second}. A final \qty{40}{\nano\second} equilibration phase was then performed in the NVT ensemble.

All production simulations were carried out in the NVT ensemble using simulation boxes with dimensions summarized in
Table~\ref{tab:systems}.

\subsection{Adaptive Biasing Force Simulations}

ABF simulations were performed with the Colvars library in NAMD~3.0.1.\cite{fiorin_using_2013,ComerGumbart2015}
The CV used for biased sampling and subsequent residence-time analysis was defined as the position of the solute
center of mass along the membrane normal ($z$), measured relative to the center of the membrane or slab, as
illustrated in Figure~\ref{fig:model_systems}. The membrane or slab remained continuous along $z$, which is the
only coordinate used in the one-dimensional transport analysis.

Mean forces were accumulated in \SI{0.1}{\angstrom}-wide bins along the CV. Biasing forces were introduced
gradually using a linear ramp from 0 to 1 until 1000 force samples had been collected in each bin. Harmonic boundary
potentials with a force constant of \SI{5}{\kcal\per\mol\per\angstrom\squared} were applied at the boundaries of
the sampled CV interval. These boundaries were placed at the ends of the bulk aqueous regions, so that the solute
remained within the primary simulation cell along $z$ and periodic-boundary crossings did not introduce ambiguities
in the center-of-mass coordinate.

For the hexadecane/water and POPC/water systems, the ABF bias was accumulated in a single continuous window
spanning the full CV interval. For the SC membrane systems, the CV was stratified into seven overlapping ABF
windows to enable parallel sampling across the membrane, following the scheme previously used by Schow
\textit{et al.},\cite{SchowFreites2011} with adjacent windows overlapping by \SI{5}{\angstrom}. Details of the
window definitions are provided in Section~S3.1.

The sufficient convergence of the ABF bias required for the RTA was assessed operationally from the time evolution of the PMF. Once the PMF became stationary and symmetric within statistical uncertainty, the residual drift was taken to
be negligible on the spatial scale of the residence-time intervals used in the analysis.

\subsection{Harmonically Restrained Simulations}

For comparison with established equilibrium fluctuation-based diffusivity estimators, we report
position-dependent diffusivities obtained from our previous study for the POPC and SC membrane
systems.\cite{ThomasPrabhakar2026} In that work, the solute was harmonically restrained at
fixed positions along the CV to generate trajectories suitable for fluctuation-based analysis.

Position-dependent diffusivities were determined using two commonly employed estimators based on the velocity and
position autocorrelation functions (VACF and PACF).\cite{WoolfRoux1994,Hummer2005}

In the VACF-based approach of Woolf and Roux,\cite{WoolfRoux1994} the velocity autocorrelation function
\begin{equation}
C_v(t)=\langle \dot{z}(t)\dot{z}(0)\rangle
\end{equation}
is used to construct a Laplace-frequency-dependent diffusivity
\begin{equation}
D(s)=
\frac{-\hat{C}_v(s)\langle \delta z^2\rangle\langle \dot{z}^2\rangle}
{\hat{C}_v(s)\!\left[s\langle \delta z^2\rangle+s^{-1}\langle \dot{z}^2\rangle\right]
-\langle \delta z^2\rangle\langle \dot{z}^2\rangle},
\end{equation}
which is extrapolated to zero frequency,
\begin{equation}
D(z=\langle z\rangle)=\lim_{s\to0}D(s).
\end{equation}

In the PACF-based approach introduced by Hummer,\cite{Hummer2005} the normalized position autocorrelation function
\begin{equation}
C_{\delta z}(t)=\frac{\langle \delta z(t)\delta z(0)\rangle}{\langle \delta z^2\rangle}
\end{equation}
yields
\begin{equation}
D(z=\langle z\rangle)=\frac{\langle \delta z^2\rangle}{\int_0^\infty C_{\delta z}(t)\,dt}.
\end{equation}

Full simulation and analysis details for the restrained trajectories are provided in
Ref.~\cite{ThomasPrabhakar2026}.

\subsection{Residence-Time Analysis}

Mean residence times were estimated from the trajectory time series $z(t)$ by identifying contiguous trajectory
segments that remained within a given interval $\Omega_i$ until first exit.

For a segment containing $m$ consecutive frames inside interval $\Omega_i$, each frame was treated as a possible
starting point and assigned the remaining time until first exit. The associated exit times are therefore
\[
m\Delta t,(m-1)\Delta t,\ldots,\Delta t,
\]
where $\Delta t$ is the trajectory sampling interval.

The interval-averaged residence time was estimated as
\begin{equation}
\widehat{\tau}_{r,i}
=
\frac{1}{N_i}\sum_{k=1}^{N_i}T_{i,k},
\end{equation}
where $T_{i,k}$ is the first-exit time associated with the $k$-th starting frame in interval $\Omega_i$, and
$N_i$ is the total number of starting frames assigned to that interval. Local diffusivities were then obtained from
the residence-time estimator derived in Theory, Eq.~\eqref{eq:rta_local_estimator}.

An interval width $L=\SI{7.5}{\angstrom}$ was used for the residence-time analysis in all systems. This choice reflects the inherent coarse-graining of the RTA: $L$ must be sufficiently large for the first-passage
statistics within an interval to be well described by an effective diffusive model, yet sufficiently small to
resolve spatial variations in the transport properties. To assess the sensitivity of the results to this choice, we analyze the dependence of the estimated diffusivities on $L$ for representative regions of the SC/water system in Section~S5.3. The observed plateau over a finite range of $L$ indicates that the reported diffusivities are robust with respect to the precise choice of domain width within this regime. The broader implications of the choice of $L$ are discussed further in the Conclusions and Outlook section.

\subsection{Uncertainty Estimation}

Because the residence times extracted from a continuous MD trajectory are temporally correlated, statistical
uncertainties cannot be obtained by assuming independent samples. We therefore estimated uncertainties using the
blocking transformation method of Flyvbjerg and Petersen,\cite{FlyvbjergPetersen1989} which determines the
asymptotic variance without assuming an explicit correlation model.

Specifically, we employed the automated blocking procedure of
Jonsson,\cite{Jonsson2018} which identifies the asymptotic variance regime and estimates the effective number of
independent samples. Uncertainties in $\widehat{\tau}_{r,i}$ were propagated to $\widehat{D}_i$ by standard error
propagation. All reported uncertainty intervals correspond to 95\% confidence intervals.

As an additional consistency check, we compared the automated blocking procedure with a variance-normalization approach for correlated data.\cite{McCluskey2025AccurateEstimation} Both methods yield closely similar estimates of the asymptotic variance for the residence-time data considered here; details are provided in Section~S5.2.

\subsection{Propagator Analysis}

To assess the predictive accuracy of the reduced transport models, we compared propagators predicted from the
Smoluchowski equation with propagators obtained directly from unbiased MD simulations.

Given a PMF $F(z)$ and a position-dependent diffusivity $D(z)$, the time evolution of the probability density
$p(z,t)$ is governed by Eq.~\eqref{eq:smoluchowski}. For a delta-function initial condition
$p(z,t_0)=\delta(z-z_0)$, the solution yields the propagator $p(z,t\,|\,z_0,t_0)$.

Reference propagators obtained directly from unbiased MD simulations for the POPC and SC systems were previously
reported in Ref.~\cite{ThomasPrabhakar2026}. In that work, initial configurations were selected from umbrella-sampling
trajectories for which the solute position satisfied $|z-z_0|<\SI{0.05}{\angstrom}$, providing a narrow initial
distribution that approximates the delta-function condition. From each
selected configuration, the umbrella restraint was then removed and the system was propagated without bias. Histograms of the solute position were accumulated to estimate
conditional probability distributions $p(z,t\,|\,z_0)$ at different lag times.

Model propagators were obtained by numerically solving the Smoluchowski equation using a forward-in-time,
centered-in-space finite-difference scheme as described in Ref.~\cite{ThomasPrabhakar2026}. Because the
diffusivity estimators yield discrete values at sampled positions, the corresponding $D(z)$ profiles were
represented by smoothing splines before solving the Smoluchowski equation. The spline interpolation employs zero
first- and second-derivative boundary conditions at the ends of the sampled interval, providing a smooth
representation of the discrete diffusivity profile for the numerical solution.

\subsection{Permeability Calculations}

Permeability coefficients for the POPC and SC membrane systems were computed from the PMF and diffusivity profiles
using the inhomogeneous solubility--diffusion model, Eq.~\eqref{eq:isd}. For each
system, the integration bounds $z_1$ and $z_2$ were determined from the corresponding density profiles. The specific
boundary definitions and numerical permeability coefficients are reported in Section~S4.

\section{Results and Discussion}

\subsection{Oxygen Diffusion in a Hexadecane/Water Slab}

We first consider oxygen diffusion across a hexadecane/water slab, which serves as a simple heterogeneous reference
system with well-defined bulk regions (Section~S1.2). This system provides the most direct test of the
residence-time approach because independently determined bulk diffusivities are available in both phases.

\begin{figure}
    \centering
    \includegraphics[width=\columnwidth]{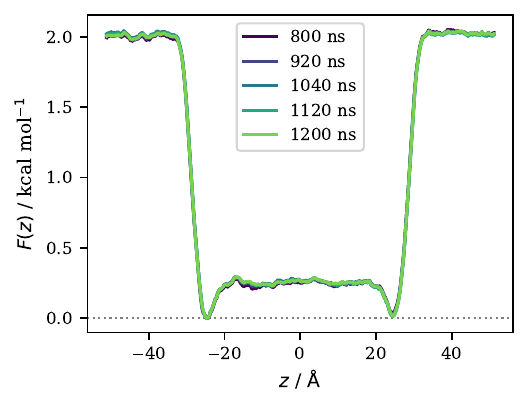}
    \caption{Convergence of PMF profiles from ABF simulations of oxygen diffusion across the hexadecane/water slab.}
    \label{fig:o2_pmf_convergence}
\end{figure}

Applying the RTA requires trajectory segments for which the effective PMF is approximately flat.
Convergence of the ABF bias was assessed from the time evolution of the PMF (Figure~\ref{fig:o2_pmf_convergence}).
From \qty{800}{\nano\second} onward, the PMF remains stationary and symmetric within statistical uncertainty,
indicating that the ABF bias has converged sufficiently for residence-time analysis. All RTA results reported below
were therefore obtained from this production interval.

The free energy difference between the bulk water and hexadecane phases is consistent with the value reported by
Ghysels \textit{et al.}\cite{GhyselsVenable2017}, indicating that the thermodynamic driving force for partitioning
is well reproduced.

\begin{figure}
    \includegraphics[width=\columnwidth]{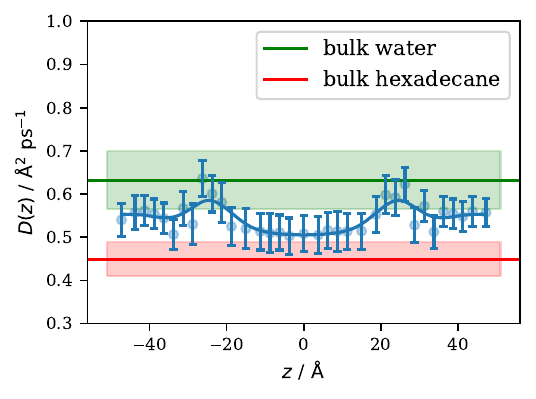}
    \caption{Diffusivity profiles for oxygen permeation across the hexadecane/water slab. Symbols denote computed
    diffusivities; lines are guides to the eye. The plotted diffusivities were obtained by averaging over 10
    individually tracked oxygen molecules. Details of the bulk diffusivity calculations and individual profile data are
    provided in Sections~S1.3 and~S1.6.}
    \label{fig::O2-diff}
\end{figure}

The diffusivity profile obtained from the RTA is shown in Figure~\ref{fig::O2-diff}, together with bulk reference
values for water and hexadecane obtained from mean-squared-displacement (MSD) analysis (Section~S1.3). Clear
plateau regions are observed in both the aqueous and hydrophobic phases, consistent with approximately constant
diffusivities in the bulk regions.

The RTA-derived bulk diffusivities are statistically compatible with the MSD reference values. Small systematic
deviations are observed, but the corresponding uncertainty intervals overlap, and the absolute differences are below
\SI{0.1}{\angstrom\squared\per\pico\second}. These discrepancies are small on the scale of the diffusivities considered here and are most likely attributable to the finite interval width used in the residence-time analysis. The agreement with independently determined bulk diffusivities therefore provides a direct benchmark for the RTA in a system where reliable reference values are accessible.

For comparison, we also analyzed the smaller hexadecane/water slab system described in Section~S1.4, which more
closely matches the setup of Ghysels \textit{et al.}\cite{GhyselsVenable2017} In that case, the reduced spatial extent of the bulk
regions leads to less well-defined plateau behavior in the diffusivity profile, consistent with the expectation
that broader bulk domains improve the robustness of local diffusivity estimates.

Overall, these results show that the RTA recovers the expected bulk limiting behavior and yields diffusivity
profiles consistent with independently determined diffusion coefficients in both phases. The hexadecane/water slab
therefore establishes a useful baseline before turning to membrane systems, in which validation relies on comparison
with independent dynamical observables rather than on direct bulk references.

\subsection{Water Permeation Across a POPC Lipid Bilayer}

We next consider water permeation across a POPC lipid bilayer, a standard benchmark for transport in fluid-phase membranes. In contrast to the slab system, direct bulk reference diffusivities no longer provide a meaningful test of the full position-dependent diffusivity profile through the membrane interior. We therefore compare the RTA with established fluctuation-based estimators and subsequently assess the resulting diffusivity profiles through propagator-level comparison with unbiased MD simulations.

\begin{figure}
    \includegraphics[width=\columnwidth]{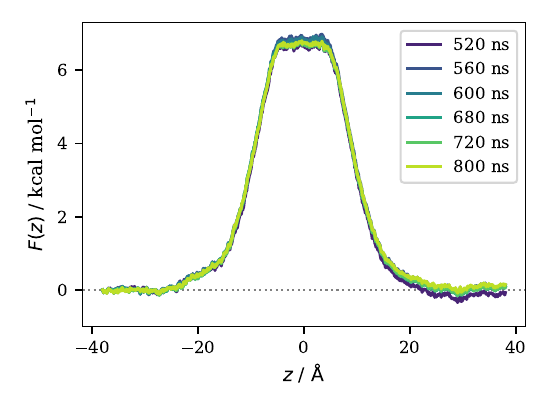}
    \caption{Convergence of PMF profiles from ABF simulations of water permeation across a POPC bilayer.}
    \label{fig:popc_fe_convergence}
\end{figure}

The validity of the RTA again requires trajectory segments for which the effective PMF is approximately flat.
Convergence of the ABF bias was assessed from the time evolution of the PMF
(Figure~\ref{fig:popc_fe_convergence}). From \qty{520}{\nano\second} onward, the PMF remains stationary and
symmetric within statistical uncertainty, indicating that the bias has converged sufficiently for residence-time
analysis. The final ABF-derived PMF is also in good agreement with the independent US estimate of the same
underlying free-energy profile reported in Ref.~\cite{ThomasPrabhakar2025}.

\begin{figure}
    \includegraphics[width=\columnwidth]{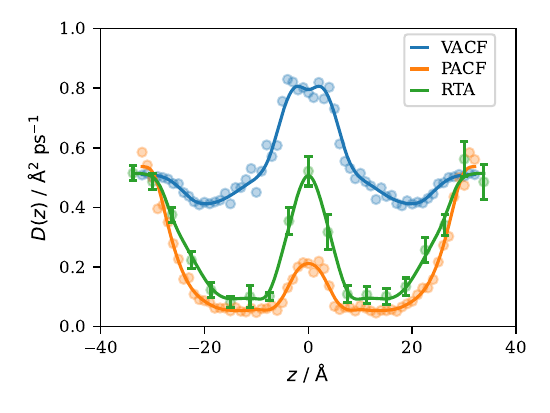}
    \caption{Diffusivity profiles across the POPC lipid bilayer obtained from the VACF, PACF, and residence-time approaches (RTA). Symbols denote the computed diffusivities, and solid lines show the smoothing-spline representations used for the propagator calculations.}
    \label{fig:popc_diffs}
\end{figure}

The diffusivity profile obtained from the RTA is shown in Figure~\ref{fig:popc_diffs}, together with profiles obtained using the VACF and PACF approaches.\cite{ThomasPrabhakar2026} All three methods yield similar diffusivities in the bulk water, whereas larger differences emerge in the bilayer interior. In particular, the RTA-derived diffusivity at the bilayer center lies between the corresponding VACF and PACF estimates.

The substantial differences between the VACF- and PACF-derived diffusivities in the membrane interior have previously been reported for water permeation across lipid bilayers.\cite{AwoonorWilliamsRowley2016,ThomasPrabhakar2026} Recent work has demonstrated that external confinement and constraints can modify the effective friction and memory kernel experienced by a solute, leading to diffusivities that differ systematically from those obtained for freely diffusing particles.\cite{Daldrop2017ExternalPotential,Yamaguchi2021DecouplingSolvent} Motivated by these observations, we conjecture that the discrepancies between the VACF and PACF estimators may arise from confinement-induced memory effects, spatial variations of the local diffusivity within the biased region, or coupling to slow solvent and membrane relaxation processes. While this interpretation remains speculative, it suggests that agreement between different equilibrium estimators cannot by itself establish the validity of a reduced diffusive description.

To assess the resulting reduced descriptions, we compared model propagators obtained by solving the Smoluchowski equation~\eqref{eq:smoluchowski} with propagators derived directly from unbiased MD simulations.\cite{ThomasPrabhakar2026} This propagator-level comparison is particularly sensitive to possible errors arising from confinement-induced changes in effective friction or memory, which may affect different diffusivity estimators in different ways.\cite{Daldrop2017ExternalPotential,Yamaguchi2021DecouplingSolvent}

\begin{figure}
    \centering
    \includegraphics[width=\columnwidth]{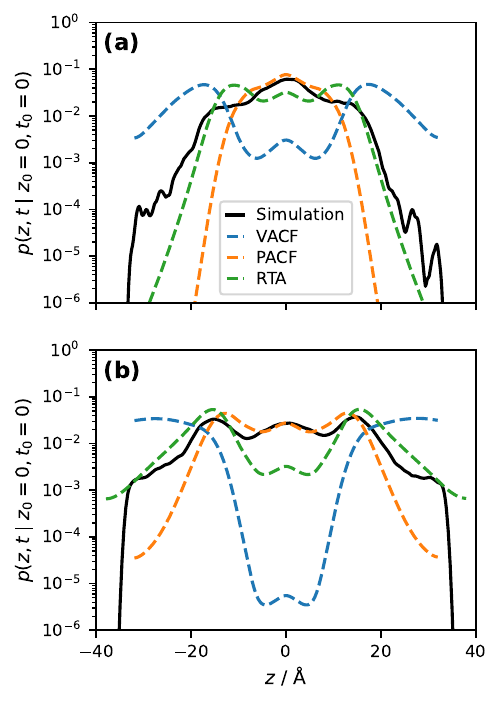}
    \caption{Comparison of MD-derived propagators with model predictions obtained using diffusivity profiles from the
    VACF, PACF, and residence-time approaches (RTA) for water permeation across the POPC bilayer. Propagators originate from $z_0=0$, corresponding to the
    membrane center. Results are shown for lag times of (a) \SI{75}{\pico\second} and
    (b) \SI{200}{\pico\second}.}
    \label{fig:popc_propagators}
\end{figure}

We focus on propagators originating from $z_0=0$, corresponding to the membrane center, where the largest
differences between the diffusivity profiles are observed (Figure~\ref{fig:popc_diffs}). Figure~\ref{fig:popc_propagators}
compares predictions obtained with each diffusivity profile against the MD-derived propagators. Results are shown for representative lag times of
\SI{75}{\pico\second} and \SI{200}{\pico\second}, which lie well within the diffusive regime. Repeating the
comparison with the US-derived PMF estimate does not change the relative performance of the three diffusivity
profiles (Section~S2.3).

All three diffusivity profiles reproduce the overall broadening of the distributions with increasing lag time.
At very short lag times up to about \SI{10}{\pico\second}, the VACF-derived diffusivity yields the closest agreement with the MD-derived propagators. However, this regime is not expected to satisfy the diffusive,
Markovian assumptions underlying a Smoluchowski description. At the representative lag times shown here, the
RTA-based diffusivity yields the closest agreement with the MD-derived propagators in the membrane center,
whereas the VACF- and PACF-based diffusivities show more pronounced deviations.

Over an extended range of lag times (Section~S2.3), the VACF-derived diffusivity systematically overestimates the diffusive spread. Neither the PACF- nor the RTA-derived diffusivity reproduces the MD propagators over the entire lag-time range considered, indicating that no single lag-time-independent diffusivity profile fully captures the projected dynamics in POPC. Nevertheless, the model based on the RTA-derived diffusivity remains in close agreement with the unbiased MD propagators over a substantial range of lag times extending to approximately \SI{200}{\pico\second}. Notably, this timescale considerably exceeds the mean residence time within the interval used to estimate the diffusivity, demonstrating that the effective diffusivity inferred from local first-exit statistics remains predictive well beyond the temporal scale from which it is derived. At the largest lag times considered, the PACF-derived diffusivity provides somewhat better agreement with the MD propagators. The observed crossover further suggests that no single Markovian diffusion model captures the projected dynamics across all timescales, consistent with previously reported subdiffusive behavior for small-molecule permeation across POPC bilayers.\cite{ChipotComer2016}

More generally, the agreement between the RTA-derived propagators and unbiased MD supports the spatial coarse-graining implied by the chosen interval width $L=\SI{7.5}{\angstrom}$, indicating that the dynamics are well described by diffusion with locally constant drift and diffusivity on this length scale. As discussed further below and in Section~S5.3, the estimated diffusivities exhibit a plateau over a finite range of $L$, indicating that the reported results are not strongly affected by the precise choice of interval width within this regime.

\subsection{Water and VOC Permeation Across the SC Lipid Membrane}

The SC membrane provides a substantially more demanding test of the RTA than POPC because of its highly ordered, multicomponent lipid organization. Unlike fluid-phase phospholipid bilayers, the SC lipid matrix consists of a heterogeneous mixture of ceramides, free fatty acids, and cholesterol that exhibits reduced molecular mobility and slow structural relaxation.\cite{ThomasPrabhakar2025} These characteristics give rise to heterogeneous transport pathways, thereby challenging reduced one-dimensional diffusion models and rendering free-energy calculations exceptionally difficult to converge.\cite{ThomasPrabhakar2025,ThomasPrabhakar2026}

To account for these slow relaxation processes, we employed a windowed ABF sampling scheme along the CV. The
convergence behavior in each window is analyzed in detail in Section~S3.2, and the RTA was applied only to
trajectory segments from the converged portions of the individual windows. Comparison with independent US-derived
PMF estimates (Section~S3.3) shows close agreement for water, whereas noticeable differences remain for the more
complex organic solutes.\cite{ThomasPrabhakar2025} Because permeability weights the barrier regions exponentially,
these residual differences can produce substantial differences in quantitative permeability estimates, as examined
below. In the propagator analysis, we hold the
ABF-derived PMF fixed across the three diffusivity profiles and assess the resulting reduced Smoluchowski models
against unbiased MD simulations. Corresponding calculations using the US-derived PMF estimate are reported in
Section~S3.4.

\begin{figure}
    \centering
    \includegraphics[width=\columnwidth]{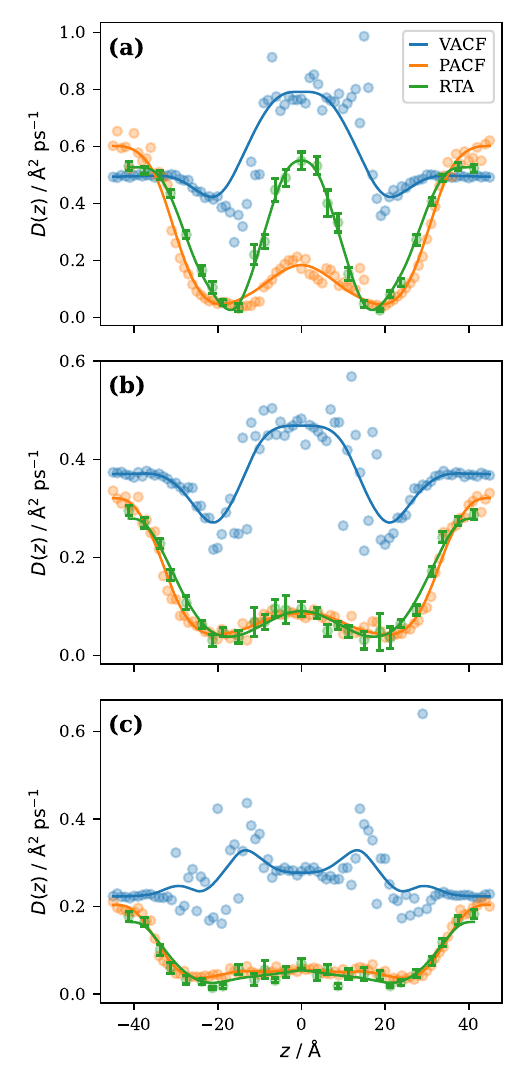}
    \caption{Position-dependent diffusivity profiles across the SC membrane for (a) water, (b) acetone, and (c) 6-MHO. Symbols denote the computed diffusivities, and solid lines show the smoothing-spline representations used for the propagator calculations.}
    \label{fig:sc_diff}
\end{figure}

The diffusivity profiles obtained from the RTA are shown in Figure~\ref{fig:sc_diff}. For water, the RTA-derived
diffusivity follows the same qualitative trend observed in POPC, lying between the VACF and PACF estimates in the
membrane center. For the organic solutes, acetone and 6-MHO, the RTA and PACF profiles are in close agreement
throughout both the membrane and bulk water regions, whereas the VACF approach yields systematically different
profiles, consistent with the trends already observed in POPC.

\begin{figure}
    \centering
    \includegraphics[width=\columnwidth]{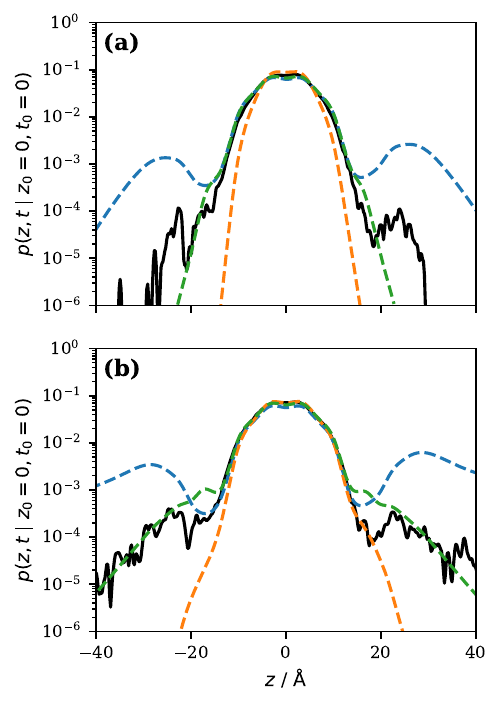}
    \caption{Comparison of MD-derived propagators with model predictions for water permeation across the SC membrane,
    obtained using diffusivity profiles from the VACF, PACF, and residence-time approaches (RTA). The common
    ABF-derived PMF was used for all model predictions. Propagators originate from $z_0=0$, corresponding to the
    membrane center. Results are shown for lag times of (a) \SI{75}{\pico\second} and
    (b) \SI{200}{\pico\second}.}
    \label{fig:sc_propagators}
\end{figure}

We applied the same propagator-level validation to the SC membrane. Figure~\ref{fig:sc_propagators} shows representative results for water at lag times of \SI{75}{\pico\second} and \SI{200}{\pico\second}. In the membrane center, the RTA-derived
diffusivity yields the closest agreement with the MD-derived propagators. In contrast to POPC, this agreement
persists over the full range of lag times considered, up to \SI{700}{\pico\second} (Section~S3.4). The relative
performance of the three diffusivity profiles is unchanged when the US-derived PMF estimate is used instead of the
ABF-derived estimate.

A similar pattern is observed for the organic solutes. For acetone and 6-MHO, the RTA and PACF approaches perform
comparably, consistent with the near-identical diffusivity profiles obtained from these two methods, whereas VACF-based propagators again predict systematically broader spreading.

Taken together, these results indicate that the RTA can provide an effective reduced description of projected transport in the SC membrane despite its structural heterogeneity and slow relaxation processes. In this system, the
residence-time estimator performs well for all solutes examined and, for water, yields the closest propagator-level agreement among the methods compared here.

\subsection{Permeabilities}

\begin{figure*}
    \centering
    \includegraphics[width=\textwidth]{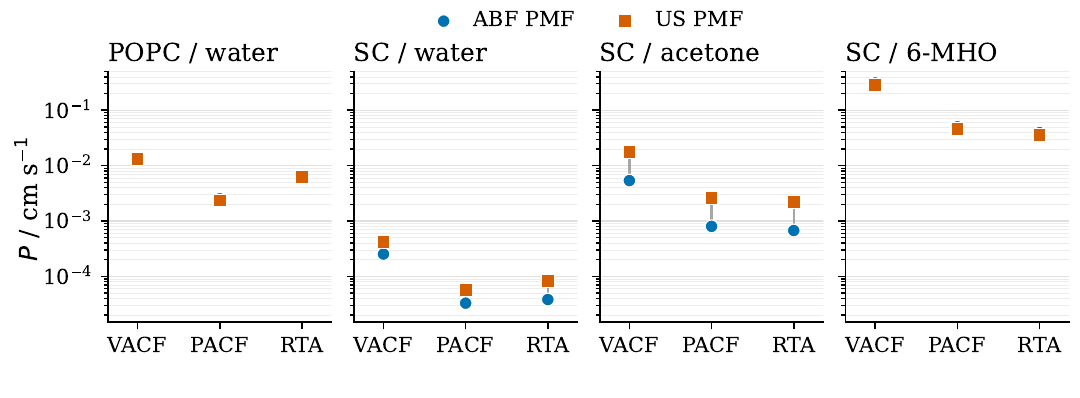}
    \caption{Sensitivity of the calculated permeability coefficients to the PMF estimate and diffusivity profile used in the
    inhomogeneous solubility--diffusion model. For each system and diffusivity estimator, permeabilities were
    calculated using both the ABF- and US-derived PMF estimates. Points connected by a vertical line use the same
    diffusivity profile and differ only in the PMF estimate; the connecting lines indicate PMF sensitivity and do not
    represent statistical uncertainty.}
    \label{fig:permeability}
\end{figure*}

To separate the effects of the PMF and diffusivity profile, we calculated permeabilities using every combination of the ABF- and US-derived PMF estimates with the RTA-, PACF-, and VACF-derived diffusivities. The results are shown in Figure~\ref{fig:permeability}, with numerical values reported in Section~S4.2.

The absolute permeabilities depend appreciably on the PMF estimate, particularly for the SC systems, where the ABF- and US-derived estimates differ in the barrier regions (Section~S3.3). Nevertheless, the relative ordering of the three diffusivity estimators is unchanged by the choice of PMF estimate: the VACF-based estimates are consistently the largest, whereas the ordering of the RTA- and PACF-based estimates is system dependent. Thus, the qualitative comparison between diffusivity estimators is robust to the PMF estimate used, while accurate quantitative permeability prediction requires well-converged PMF and diffusivity profiles.

\section{Conclusions and Outlook}

We have introduced a residence-time approach (RTA) for determining position-dependent diffusivities from biased molecular simulations in the drift-free limit. The method uses trajectory segments for which the effective free-energy gradient along the transport coordinate is negligible and relates mean first-exit times from finite spatial intervals directly to effective local diffusivities. In the present work, this regime was realized using converged ABF simulations, but the formulation is not specific to ABF and can in principle be combined with other biasing strategies that generate an approximately flat effective free-energy landscape.

The RTA provides a practical alternative to fluctuation-based estimators based on harmonically restrained simulations. By extracting diffusivities from residence-time statistics in an approximately drift-free landscape, it avoids inferring transport coefficients from equilibrium fluctuations around an imposed restraint, a setting in which external confinement may alter the effective friction and memory kernel experienced by the solute.\cite{Daldrop2017ExternalPotential,Yamaguchi2021DecouplingSolvent} It also bypasses the correlation-function integrations required by conventional VACF- and PACF-based estimators.\cite{WoolfRoux1994,Hummer2005,GaalswykAwoonorWilliams2016}

We evaluated the method from a simple liquid--liquid reference system with independently accessible bulk diffusivities to increasingly heterogeneous membrane systems where validation must rely on dynamical consistency. In the hexadecane/water system, where independent bulk reference diffusivities are available, the RTA reproduces the bulk diffusion coefficients in both phases within statistical uncertainty, providing a direct benchmark in a simple heterogeneous reference system.

For the membrane systems, direct bulk reference diffusivities are not available across the full transport coordinate. We therefore assessed the inferred diffusivity profiles by propagator-level validation against unbiased MD simulations. The primary comparison holds the ABF-derived PMF fixed across all diffusivity profiles, while calculations using the US-derived PMF estimate show that the relative ranking of the diffusivity profiles is unchanged.

In the POPC bilayer, the different diffusivity estimators provide the closest agreement with MD-derived propagators over different lag-time regimes. VACF-derived diffusivities perform best at very short lag times, although this regime lies outside the diffusive, Markovian regime that motivates a Smoluchowski description. At intermediate lag times, the RTA gives the closest agreement with the MD-derived propagators, whereas PACF-based diffusivities perform better at the longest lag times considered here. Taken together, these results indicate that the projected POPC dynamics cannot be represented over the full lag-time range by a single lag-time-independent one-dimensional Smoluchowski model. Instead, the apparent optimal diffusivity depends on the dynamical regime being probed, consistent with residual non-Markovianity, imperfect separation of slow membrane and solvent relaxation from solute motion, and limitations inherent to projecting the full dynamics onto a single membrane-normal coordinate.

In the more ordered and heterogeneous SC membrane, the RTA performs well across all solutes examined. For acetone and 6-MHO, the RTA and PACF diffusivities are very similar throughout the membrane, whereas VACF-based estimates again predict systematically broader propagators. For water, the RTA provides the most accurate propagator-level description among the methods compared here.

Permeabilities computed from all combinations of the two PMF estimates and three diffusivity profiles separate the kinetic contribution from the sensitivity to the thermodynamic input. The relative ordering associated with the diffusivity estimators is unchanged upon substituting the US-derived PMF estimate for the ABF-derived estimate, but the absolute permeability coefficients can change appreciably, particularly for the SC systems. The membrane results therefore reinforce two main conclusions: first, that the RTA yields diffusivities comparable to those from established estimators and, in several cases, gives closer agreement with MD-derived propagators among the methods compared here; and second, that quantitative permeability prediction is highly sensitive to convergence of both the kinetic and thermodynamic inputs.

Taken together, these results support residence-time statistics as a practical route for extracting position-dependent diffusivities from biased MD simulations. At the same time, the remaining differences between model and simulation propagators indicate that the inferred diffusivities should not be interpreted as uniquely exact local transport coefficients. Rather, they should be viewed as components of an effective reduced stochastic model, whose validity is best assessed by its ability to reproduce independent dynamical observables such as propagators.

An important conclusion of the present work concerns the role of the interval width $L$. Rather than being a purely numerical parameter, $L$ should be regarded as the intrinsic spatial coarse-graining length of the RTA. In this respect, it plays a role analogous to the lag time used when extracting diffusion coefficients from mean-squared displacement analyses: no universal optimal choice can be expected, and instead one seeks an intermediate regime in which the assumptions underlying diffusive, Markovian dynamics are satisfied while preserving sufficient spatial resolution. For the systems considered here, the analysis presented in Section~S5.3 reveals a clear plateau in the estimated diffusivities over a finite range of $L$, encompassing the value used throughout this work. Together with the propagator analysis, this supports the chosen coarse-graining as a robust reduced description.

The dependence of the estimated diffusivity on $L$ reflects that residence-time statistics are evaluated over finite spatial intervals. This is analogous in spirit to the finite-window effects encountered in local displacement-based estimators, for which form-factor correction schemes have recently been proposed.\cite{Nagai2020PositiondependentDiffusion,Kikkawa2026LocalDiffusion} Although the present work does not derive an analogous correction for first-exit statistics, this connection suggests a useful direction for future developments of the RTA. A systematic investigation of the dependence on $L$ across a broader range of systems, including those exhibiting pronounced non-Markovian or anomalous dynamics,\cite{ChipotComer2016} could establish best-practice guidelines for selecting $L$ and may ultimately enable mechanistically motivated correction schemes for finite interval widths.

A second open question concerns the contrast between the POPC and SC results. Although POPC is structurally more disordered than SC, it appears to be more difficult to describe using a single lag-time-independent diffusivity profile. Determining whether this difference reflects stronger memory effects, different coupling of the chosen coordinate to orthogonal degrees of freedom, or differences in local structural relaxation will help define the range of validity of reduced one-dimensional diffusion models for membrane transport.

A third question concerns the distinct behavior of water. In both membrane systems, water shows larger differences
between RTA- and PACF-derived diffusivities than the organic solutes, whereas for acetone and 6-MHO the two
approaches yield closely similar results. This suggests that water is particularly sensitive to dynamical effects
that are captured differently by diffusivity estimators, possibly because of its small size,
hydrogen-bonding capability, and coupling to local membrane fluctuations. A broader comparison across solute classes
may therefore help identify when different estimators converge to the same effective description and when they do
not.

\section*{Data Availability}

Initial structures, simulation input files, and the software library implementing the residence-time approach are
available on GitHub at \url{https://github.com/mvondomaros-lab/rta-paper-data} and
\url{https://github.com/mvondomaros-lab/pdda}, respectively. The \texttt{pdda} library includes the
implementation of the residence-time diffusivity estimator, the associated uncertainty quantification procedures,
the smoothing-spline interpolation of discrete diffusivity profiles, and the numerical propagator analysis
described in this work. Additional data supporting the findings of this study are available from the corresponding
author upon reasonable request.

\section*{Acknowledgments}

PRP acknowledges the Wolfsberg Graduate Research Fellowship. The authors gratefully acknowledge Douglas J.
Tobias for many helpful and insightful discussions.

The authors used OpenAI's ChatGPT during the preparation of this manuscript to assist with language editing,
improving clarity, and revising the presentation of the text. All scientific content, analyses, interpretations,
and conclusions were developed by the authors, who reviewed and take full responsibility for the final manuscript.

\begin{suppinfo}
Supporting Information Available: Additional analyses of the hexadecane/water slab systems, including system definitions, orientational distributions, bulk reference diffusion coefficients, residence-time analysis of the smaller slab system, and supplementary PMF and diffusivity profiles; supplementary POPC analyses, including PMF evolution before bias convergence, comparison of ABF- and umbrella-sampling-derived PMF estimates, and propagator sensitivity to the PMF estimate; supplementary stratum corneum analyses, including ABF window definitions, window-wise PMF convergence, comparison of PMF estimates from different sampling methods, and propagator sensitivity to the PMF estimate; supplementary permeability analyses, including the definition of integration bounds and the full set of permeability coefficients; and methodological validation of the residence-time approach, including local dynamical statistics, uncertainty quantification for correlated residence-time data, and dependence on the residence interval width.
\end{suppinfo}

\bibliography{main}

\end{document}


\maketitle

\section{\ce{O2} in the Hexadecane/Water Slab}

\subsection{Hexadecane/Water Slab Systems}

In addition to the hexadecane/water slab system analyzed in the main text, hereafter denoted HS-II, we
investigated a smaller system, HS-I, with approximately half the box length in the \(z\) direction. HS-I
corresponds to the smaller slab system studied previously by Ghysels \textit{et al.}\cite{GhyselsVenable2017}

\begin{figure}[H]
    \centering
    \includegraphics[width=\textwidth]{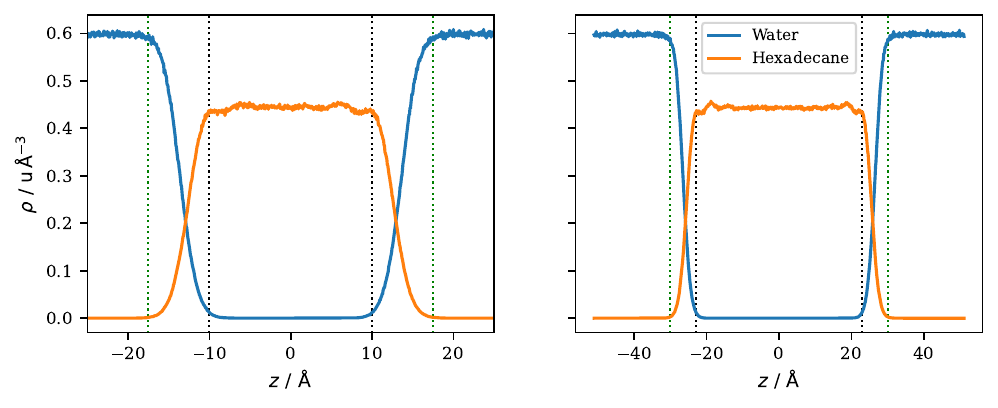}
    \caption{
        Density profiles of the two hexadecane/water slab systems considered in this work. The profiles were
        obtained from an unbiased \SI{40}{\nano\second} production run. The black and green vertical dashed lines
        indicate the boundaries of the bulk-like regions of hexadecane and water, respectively.
    }
    \label{fig:combined_density}
\end{figure}

The system compositions and box dimensions are listed in Table~\ref{tab:slab_systems}. Density profiles of the
two slab systems are shown in Figure~\ref{fig:combined_density}.

\begin{table}[H]
    \centering
    \caption{
        Compositions of the hexadecane/water slab systems and the corresponding simulation box dimensions.
    }
    \label{tab:slab_systems}
    \begin{tabular*}{\columnwidth}{@{\extracolsep{\fill}}l
        S[table-format=3.0]
        S[table-format=4.0]
        S[table-format=2.0]
        S[table-format=2.2]
        S[table-format=2.2]
        S[table-format=3.2]}
        \toprule
        System & {Hexadecane} & {Water} & {\ce{O2}} & {$L_x$ (\si{\angstrom})} &
        {$L_y$ (\si{\angstrom})} & {$L_z$ (\si{\angstrom})} \\
        \midrule
        HS-I & 126 & 2159 & 10 & 50.00 & 50.00 & 53.20 \\
        HS-II & 252 & 4318 & 10 & 50.00 & 50.00 & 105.00 \\
        \bottomrule
    \end{tabular*}
\end{table}

\subsection{Orientational Distributions}

In this section, we characterize molecular orientations in the hexadecane/water slab relative to the $z$-axis
to assess the extent of bulk-like behavior in HS-I and HS-II. The larger slab system, HS-II, was selected for
the main-text analysis because, unlike HS-I, it exhibits extended regions consistent with bulk-like behavior in
both phases.

\begin{figure}[H]
    \centering
    \includegraphics[width=0.5\textwidth]{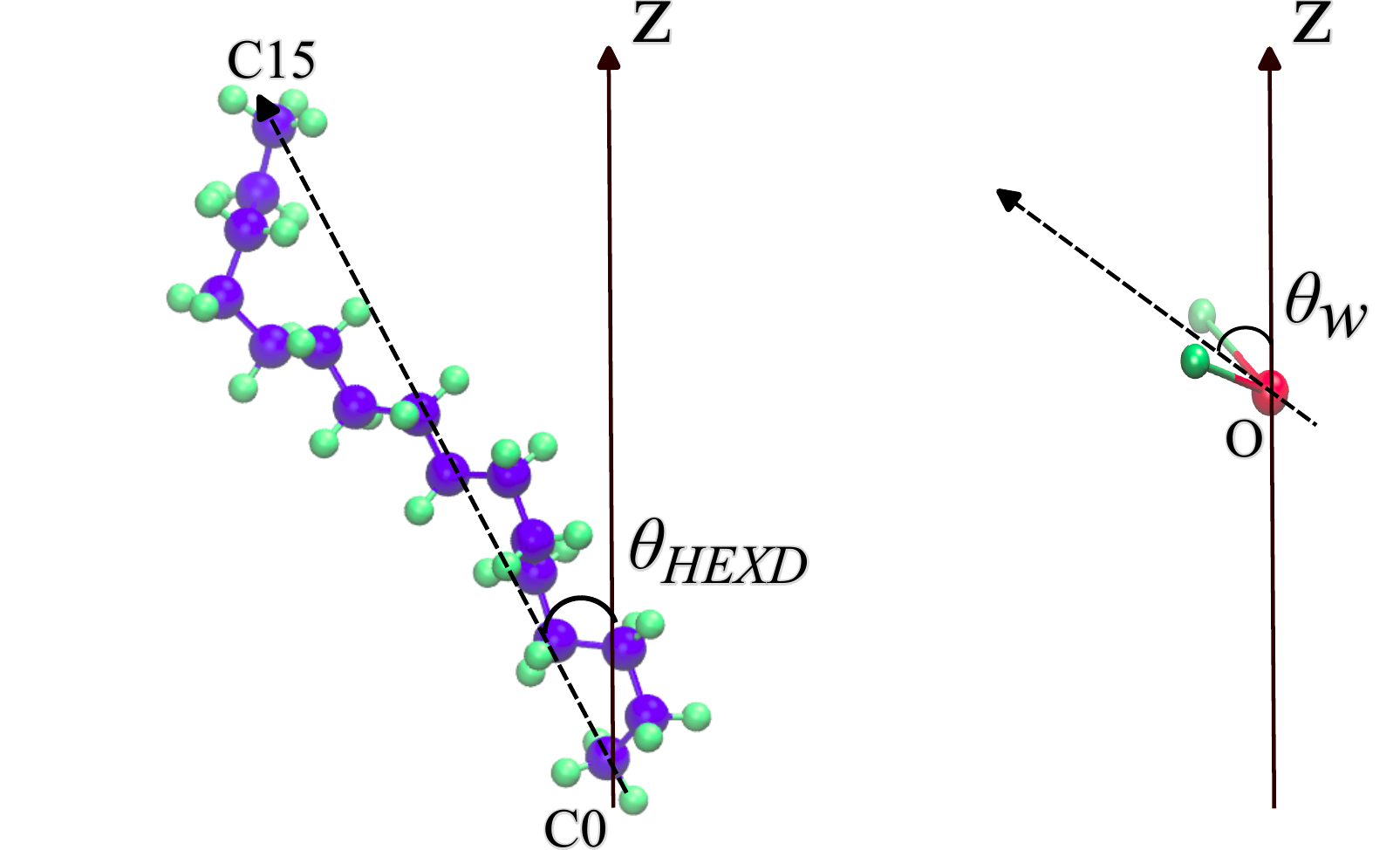}
    \caption{
        Schematic definition of the orientation angles used in the orientational analysis. Left panel:
        hexadecane; right panel: water. Carbon, hydrogen, and oxygen atoms are shown in blue, green, and red,
        respectively. The dotted line denotes the molecular orientation vector, and the vertical dashed line
        denotes the $z$-axis.
    }
    \label{fig:diagra_OP}
\end{figure}

\begin{figure}[H]
    \centering
    \includegraphics[width=\textwidth]{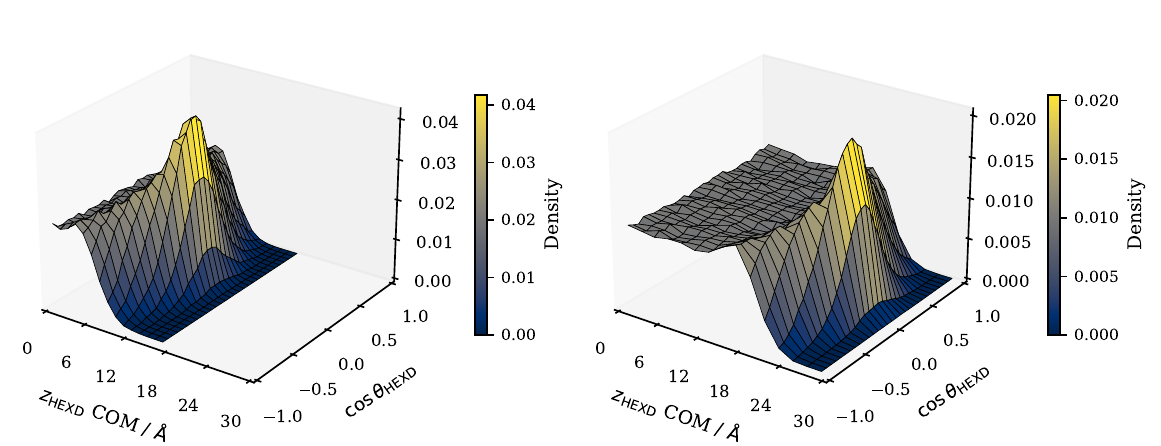}
    \caption{
        Three-dimensional distributions of $\cos \theta_{\mathrm{HEXD}}$ as a function of the hexadecane
        center-of-mass position, $z_{\mathrm{HEXD}}$, for HS-I (left) and HS-II (right).
    }
    \label{fig:hexd_op2}
\end{figure}

\begin{figure}[H]
    \centering
    \includegraphics[width=\textwidth]{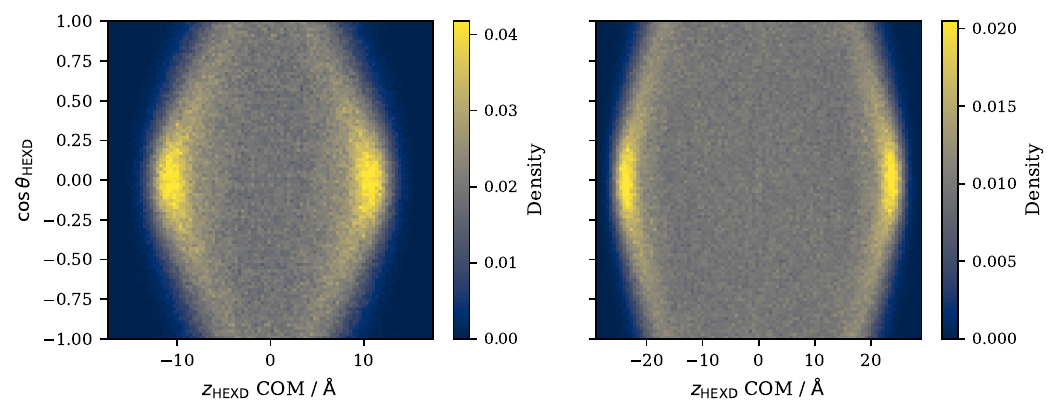}
    \caption{
        Two-dimensional distributions of $\cos \theta_{\mathrm{HEXD}}$ as a function of the hexadecane
        center-of-mass position, $z_{\mathrm{HEXD}}$, for HS-I (left) and HS-II (right).
    }
    \label{fig:hexd_op2_2d}
\end{figure}

\begin{figure}[H]
    \centering
    \includegraphics[width=\textwidth]{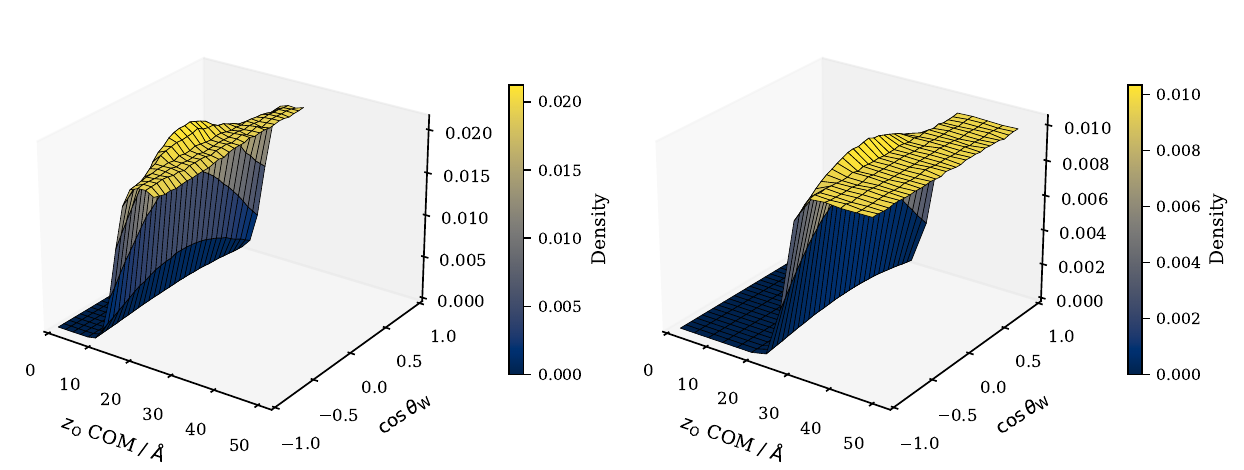}
    \caption{
        Three-dimensional distributions of $\cos \theta_{\mathrm{W}}$ as a function of the water oxygen
        position, $z_{\mathrm{O}}$, for HS-I (left) and HS-II (right).
    }
    \label{fig:water_op2}
\end{figure}

\begin{figure}[H]
    \centering
    \includegraphics[width=\textwidth]{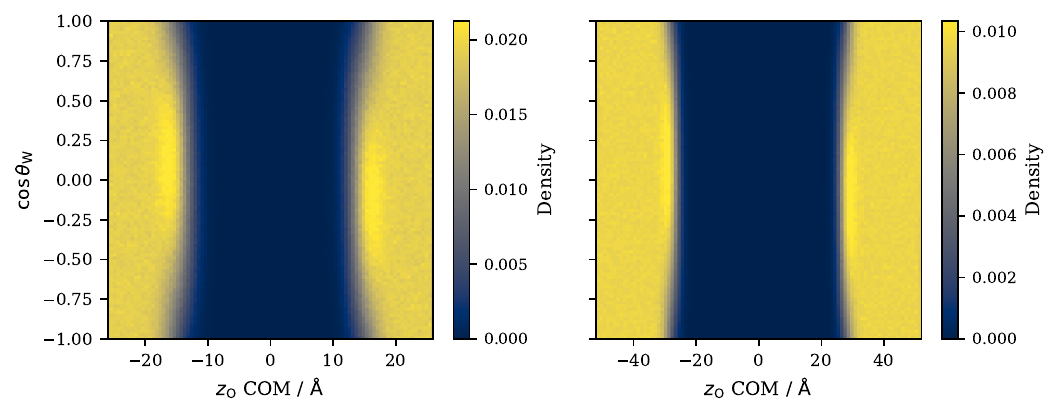}
    \caption{
        Two-dimensional distributions of $\cos \theta_{\mathrm{W}}$ as a function of the water oxygen
        position, $z_{\mathrm{O}}$, for HS-I (left) and HS-II (right).
    }
    \label{fig:water_op2_2d}
\end{figure}

For hexadecane, we defined the orientation angle, $\theta_{\mathrm{HEXD}}$, as the angle between the $z$-axis
and the vector connecting the first and terminal carbon atoms of an aliphatic chain
(Figure~\ref{fig:diagra_OP}). For water, we defined the orientation angle, $\theta_{\mathrm{W}}$, as the angle
between the $z$-axis and the molecular $C_2$ axis, defined here as the vector connecting the oxygen atom and the
midpoint between the two hydrogen atoms (Figure~\ref{fig:diagra_OP}). The orientational distributions are
expressed in terms of $\cos \theta_{\mathrm{HEXD}}$ and $\cos \theta_{\mathrm{W}}$ as functions of position
along the $z$-axis (Figures~\ref{fig:hexd_op2}--\ref{fig:water_op2_2d}). These distributions were computed
from an unbiased \SI{40}{\nano\second} production run. The distributions were symmetrized with respect to the slab
center. For hexadecane, we used $p(|\cos \theta_{\mathrm{HEXD}}|, z)=p(|\cos \theta_{\mathrm{HEXD}}|,-z)$,
reflecting the apolar, axis-like character of the molecular orientation. For water, we used
$p(\cos \theta_{\mathrm{W}}, z)=p(-\cos \theta_{\mathrm{W}}, -z)$, reflecting the inversion symmetry of the two
interfaces.

In contrast to HS-I, HS-II exhibits extended regions with only weak variation in the orientational distributions
in both the hexadecane phase (Figures~\ref{fig:hexd_op2} and \ref{fig:hexd_op2_2d}) and the water phase
(Figures~\ref{fig:water_op2} and \ref{fig:water_op2_2d}), consistent with the presence of bulk-like domains.

\subsection{Bulk Reference Diffusion Coefficients}

To determine bulk diffusion coefficients for \ce{O2} in hexadecane and water, we constructed bulk reference
systems with dimensions comparable to those of the bulk regions identified from the density profiles of the
hexadecane/water slab systems (Figure~\ref{fig:combined_density}). The bulk systems derived from HS-I are denoted
Water-I and Hexadecane-I, whereas those derived from HS-II are denoted Water-II and Hexadecane-II. Their
compositions and box dimensions are listed in Table~\ref{tab:bulk_systems}.

Initial configurations for all bulk systems were generated with Packmol. Each system was minimized for
\num{50000} steps, followed by equilibration in the NVT ensemble for \SI{4}{\nano\second}, equilibration in the
NPT ensemble for \SI{10}{\nano\second}, and a further NVT equilibration for \SI{10}{\nano\second}. Thermostat
and barostat settings were identical to those used for the slab systems described in the main text.

\begin{table}[H]
    \centering
    \caption{
        Compositions of the bulk reference systems used for the calculation of \ce{O2} diffusion coefficients
        and the corresponding simulation box dimensions.
    }
    \label{tab:bulk_systems}
    \begin{tabular*}{\columnwidth}{@{\extracolsep{\fill}}l
        S[table-format=4.0]
        S[table-format=4.0]
        S[table-format=1.0]
        S[table-format=2.2]
        S[table-format=2.2]
        S[table-format=2.2]}
        \toprule
        System & {Hexadecane} & {Water} & {\ce{O2}} & {$L_x$ (\si{\angstrom})} &
        {$L_y$ (\si{\angstrom})} & {$L_z$ (\si{\angstrom})} \\
        \midrule
        Water-I & 0 & 1418 & 1 & 50.00 & 50.00 & 17.00 \\
        Hexadecane-I & 99 & 0 & 1 & 50.00 & 50.00 & 20.00 \\
        Water-II & 0 & 3420 & 1 & 50.00 & 50.00 & 41.00 \\
        Hexadecane-II & 227 & 0 & 1 & 50.00 & 50.00 & 46.00 \\
        \bottomrule
    \end{tabular*}
\end{table}

For each bulk system, \num{10} independent production simulations of \SI{20}{\nano\second} each were carried out
in the NVT ensemble, and the center-of-mass position of \ce{O2} was recorded every \SI{20}{\femto\second}. Each
trajectory was divided into four \SI{5}{\nano\second} blocks, yielding \num{40} analysis segments in total.
Because diffusion in these bulk systems is isotropic, we computed the three-dimensional mean squared displacement
(MSD) and estimated the diffusion coefficient from the slope of the Einstein relation,
Equation~\ref{eq:einstein}, by linear regression over the interval from \SI{10}{\pico\second} to
\SI{1000}{\pico\second}.

\begin{equation}
    \lim_{t \to \infty} \left\langle \left| \vec{r}(t) - \vec{r}(0) \right|^{2} \right\rangle = 6 D t,
    \label{eq:einstein}
\end{equation}

\begin{figure}[H]
    \centering
    \includegraphics[width=\textwidth]{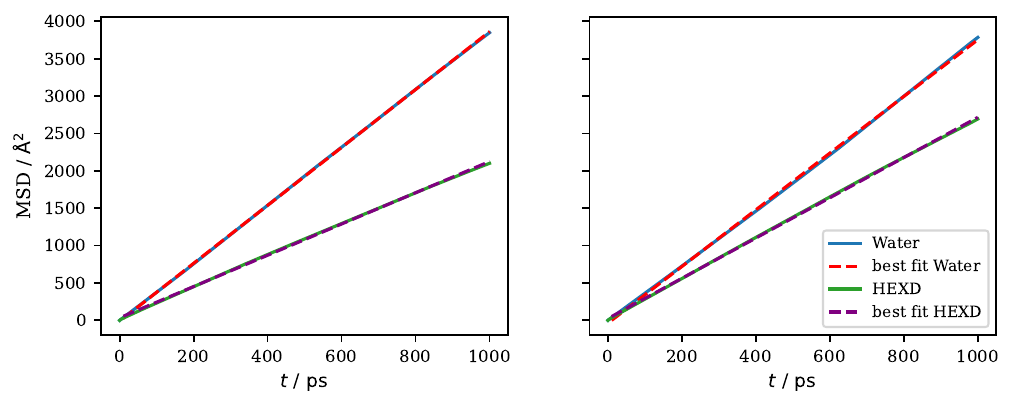}
    \caption{
        Three-dimensional mean squared displacements of \ce{O2} in the four bulk reference systems considered in
        this study. Left panel: Water-I and Hexadecane-I; right panel: Water-II and Hexadecane-II. Diffusion
        coefficients obtained from linear fits to the MSDs are listed in Table~\ref{tab:diffusion_constants}.
    }
    \label{fig:bulk_o2_msd}
\end{figure}

Figure~\ref{fig:bulk_o2_msd} shows the MSD curves and corresponding linear fits for the four bulk reference
systems. The resulting diffusion coefficients are listed in Table~\ref{tab:diffusion_constants}. The
\ce{O2} diffusion coefficient in Water-I agrees with the literature value reported by Ghysels
\textit{et al.}\cite{GhyselsVenable2017}, whereas that in Hexadecane-I is lower than the corresponding
literature value.

\begin{table}[H]
    \centering
    \caption{
        Bulk diffusion coefficients of \ce{O2} in water and hexadecane, reported in
        \si{\angstrom\squared\per\pico\second}, obtained from the Einstein relation
        (Eq.~\ref{eq:einstein}). Uncertainties were estimated from the standard error of the mean and are
        reported as \SI{95}{\percent} confidence intervals. Literature values are from
        Ghysels \textit{et al.}\cite{GhyselsVenable2017}
    }
    \label{tab:diffusion_constants}
    \begin{tabular*}{\columnwidth}{@{\extracolsep{\fill}}lcc}
        \toprule
        System & Diffusion coefficient & Literature \\
        \midrule
        Water-I & $\num{0.646} \pm \num{0.055}$ & \num{0.60} \\
        Hexadecane-I & $\num{0.349} \pm \num{0.052}$ & \num{0.44} \\
        Water-II & $\num{0.632} \pm \num{0.067}$ & --- \\
        Hexadecane-II & $\num{0.449} \pm \num{0.039}$ & --- \\
        \bottomrule
    \end{tabular*}
\end{table}

\subsection{HS-I Residence-Time Analysis}

\begin{figure}[H]
    \centering
    \includegraphics[width=0.5\textwidth]{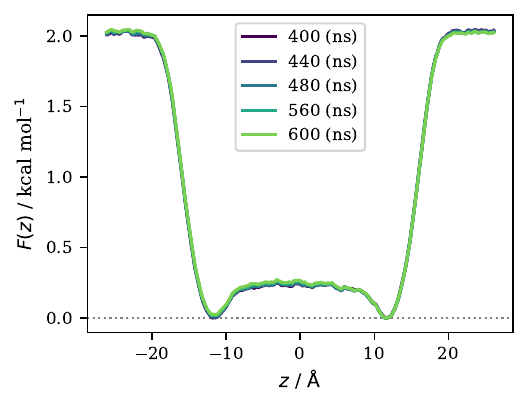}
    \caption{
        Convergence of the PMF from ABF simulations of \ce{O2} permeation across the HS-I
        hexadecane/water slab.
    }
    \label{fig:small_o2_pmf_convergence}
\end{figure}

We next applied the residence-time analysis to HS-I, the smaller hexadecane/water slab system corresponding to
the system studied by Ghysels \textit{et al.}\cite{GhyselsVenable2017} Figure~\ref{fig:small_o2_pmf_convergence}
shows the convergence of the PMF obtained from the ABF simulations. The PMF becomes approximately stationary
after \SI{400}{\nano\second}, and only data collected thereafter were used for the diffusivity analysis.

\begin{figure}[H]
    \centering
    \includegraphics[width=0.5\columnwidth]{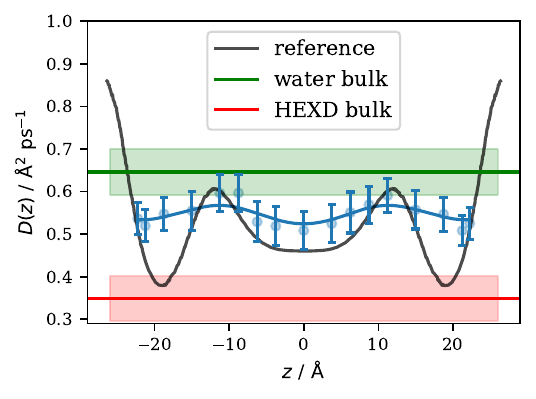}
    \caption{
        Diffusivity profile for \ce{O2} permeation across the HS-I hexadecane/water slab. Symbols denote the computed diffusivities; solid lines show spline interpolations included for visual guidance only. Error bars represent uncertainties estimated
        from blocking analysis of \num{10} independent \ce{O2} trajectories. The literature profile was digitized
        from Ghysels \textit{et al.}\cite{GhyselsVenable2017}
    }
    \label{fig:small_o2_diffusivity_final}
\end{figure}

The diffusivity profile obtained from the RTA (Figure~\ref{fig:small_o2_diffusivity_final}) differs
qualitatively from that reported by Ghysels \textit{et al.}\cite{GhyselsVenable2017} In particular, the
RTA-derived profile is comparatively flat and shows substantially less spatial structure.

We note that the RTA-derived profile reproduces the bulk-water diffusivity more closely than the
bulk-hexadecane diffusivity. At the same time, our orientational analysis indicates that fully bulk-like behavior should not necessarily be expected in the interior of the HS-I hexadecane slab. We therefore consider
two possible explanations for the qualitative differences between the present profile and that reported by
Ghysels \textit{et al.}\cite{GhyselsVenable2017} First, Ghysels \textit{et al.} represented the diffusivity
profile using a truncated Fourier cosine expansion. While this basis was introduced to smooth the profile and
avoid overfitting, the inferred profile may nevertheless retain basis-dependent structure. Second, the RTA
assumes that the diffusivity is approximately constant within each spatial interval. In a small system such as
HS-I, where bulk-like regions are limited, this approximation may smooth or attenuate genuine spatial variation
in the diffusivity profile.

At present, we do not attempt to distinguish definitively between these possibilities. However, the main
conclusions of this work for the membrane systems are unaffected, because their consistency has been corroborated
by propagator-level analysis in the main text.

\subsection{Additional HS-I Profiles}

\begin{figure}[H]
    \centering
    \includegraphics[width=\textwidth]{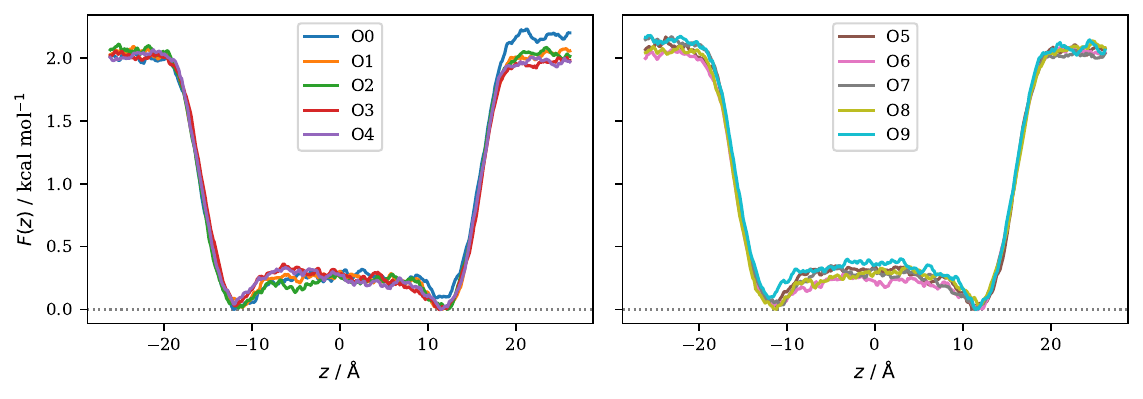}
    \caption{
        Individual raw PMF profiles for \ce{O2} permeation across the HS-I hexadecane/water slab after
        \SI{600}{\nano\second} of ABF simulation.
    }
    \label{fig:small_o2_PMF_raw}
\end{figure}

\begin{figure}[H]
    \centering
    \includegraphics[width=\textwidth]{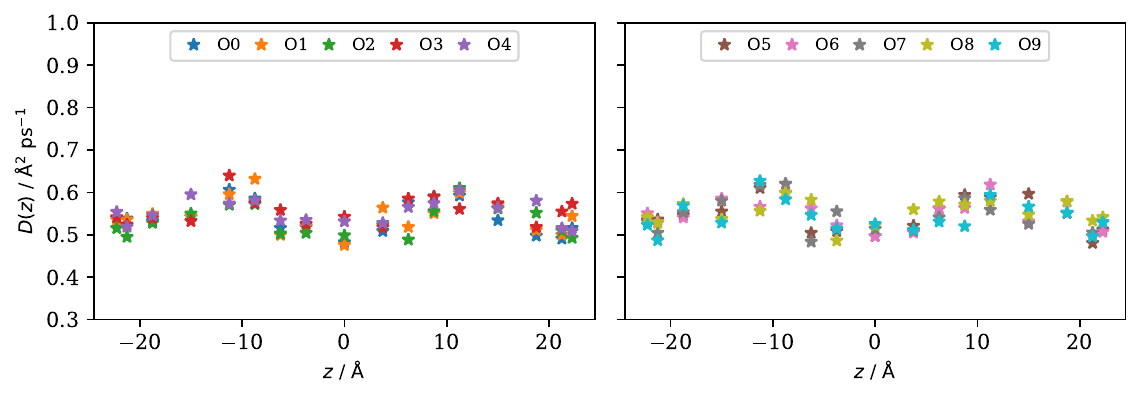}
    \caption{
        Individual raw diffusivity profiles for \ce{O2} permeation across the HS-I hexadecane/water slab.
    }
    \label{fig:small_o2_diffusivity_raw}
\end{figure}

\begin{figure}[H]
    \centering
    \includegraphics[width=\textwidth]{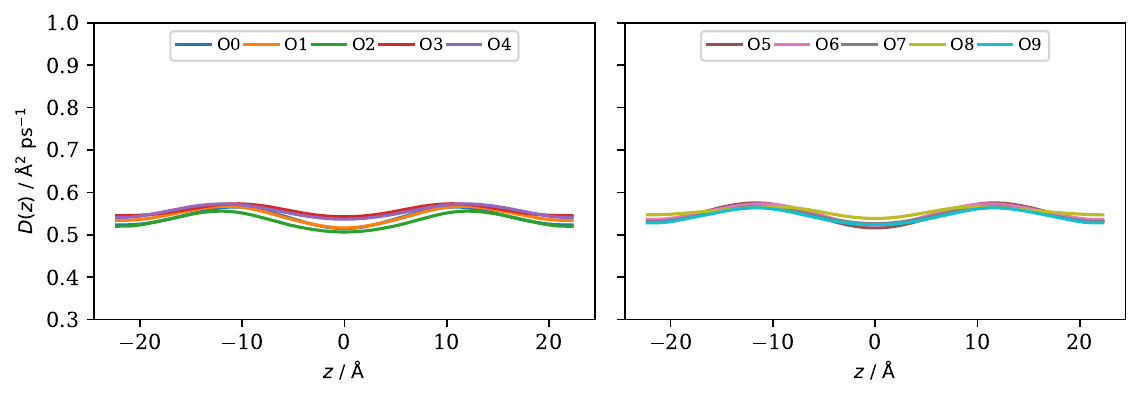}
    \caption{
        Individual spline-smoothed diffusivity profiles for \ce{O2} permeation across the HS-I
        hexadecane/water slab.
    }
    \label{fig:small_o2_diffusivity_smooth}
\end{figure}

Additional PMF and diffusivity data for HS-I are shown in
Figures~\ref{fig:small_o2_PMF_raw}--\ref{fig:small_o2_diffusivity_smooth}. These plots show the
individual profiles obtained for each \ce{O2} molecule and complement the averaged diffusivity profile shown in
Figure~\ref{fig:small_o2_diffusivity_final}.

\subsection{Additional HS-II Profiles}

\begin{figure}[H]
    \centering
    \includegraphics[width=\textwidth]{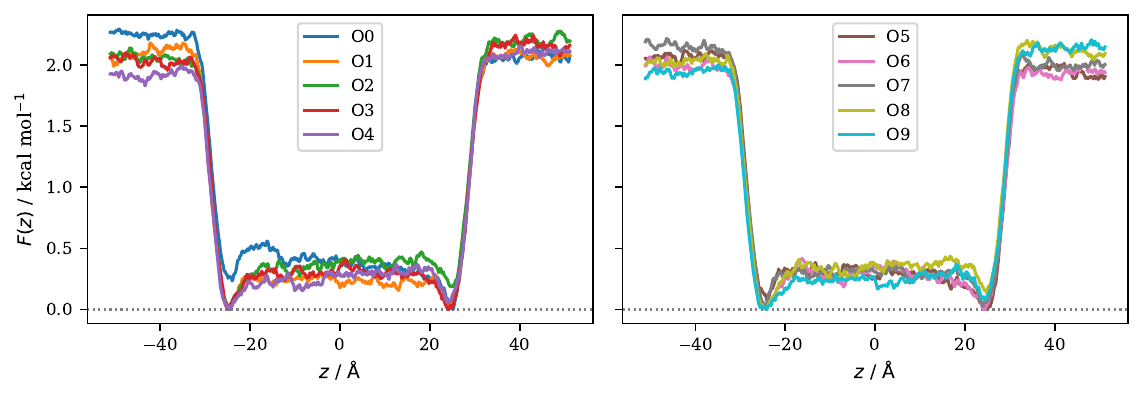}
    \caption{
        Individual PMFs for \ce{O2} permeation across the HS-II hexadecane/water slab. The
        corresponding average profile is shown in the main text.
    }
    \label{fig:big_o2_individual_pmf}
\end{figure}

\begin{figure}[H]
    \centering
    \includegraphics[width=\textwidth]{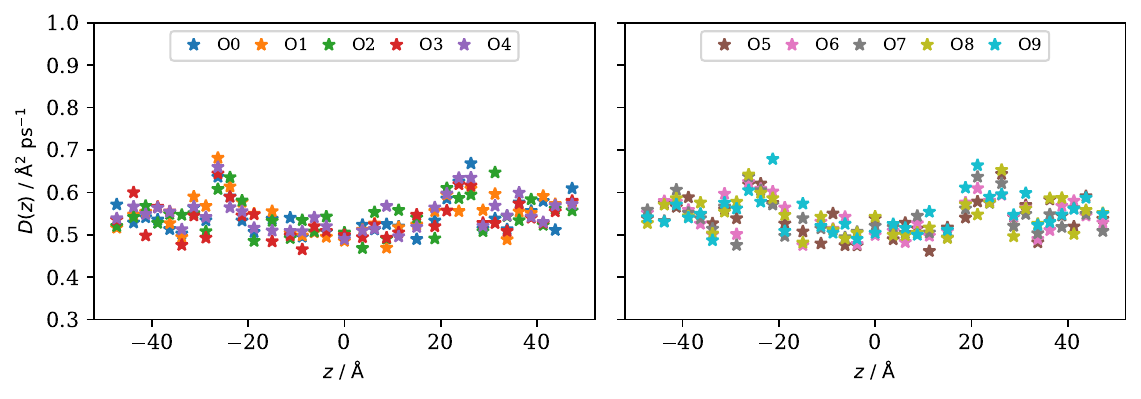}
    \caption{
        Individual raw diffusivity profiles for \ce{O2} permeation across the HS-II hexadecane/water slab.
    }
    \label{fig:big_o2_diffusivity_raw}
\end{figure}

\begin{figure}[H]
    \centering
    \includegraphics[width=\textwidth]{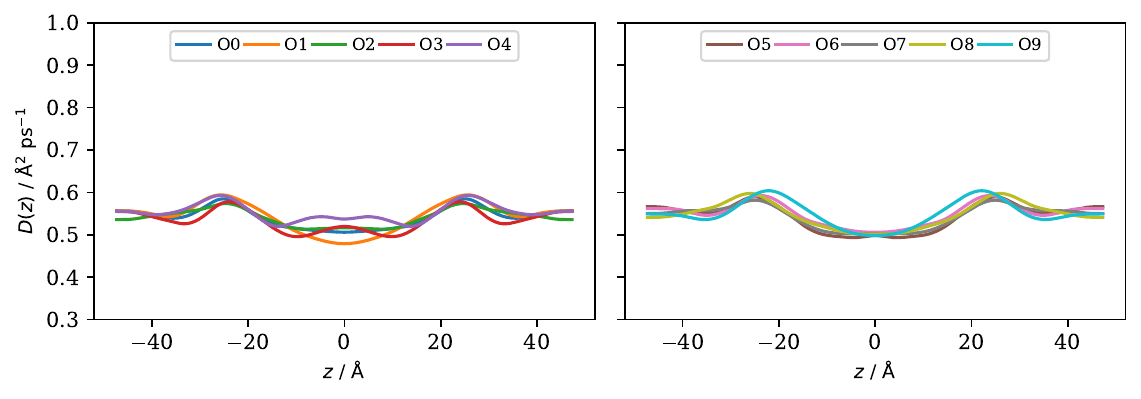}
    \caption{
        Individual spline-smoothed diffusivity profiles for \ce{O2} permeation across the HS-II
        hexadecane/water slab. The corresponding average profile is shown in the main text.
    }
    \label{fig:big_o2_diffusivity_smooth}
\end{figure}

Additional PMF and diffusivity data for HS-II are shown in
Figures~\ref{fig:big_o2_individual_pmf}--\ref{fig:big_o2_diffusivity_smooth}. These plots show the individual
profiles obtained for each \ce{O2} molecule and thus complement the averaged results presented in the main text.

\section{Water Permeation through the POPC Bilayer}

\subsection{PMF Evolution before Bias Convergence}

\begin{figure}[H]
    \centering
    \includegraphics[width=0.5\textwidth]{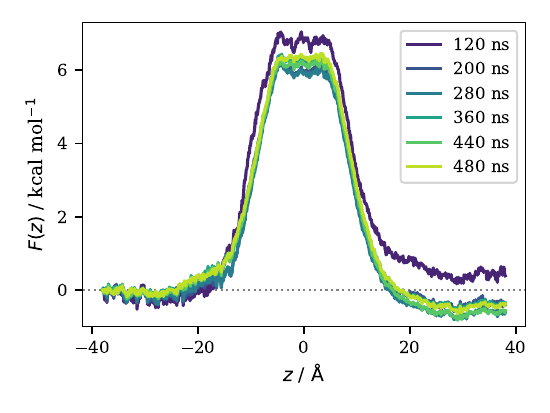}
    \caption{
        Evolution of the PMF during the initial stage of the ABF simulation, showing asymmetry between the two
        membrane leaflets. Because the ABF bias is not yet converged during this time period, these trajectory
        data were not used for the RTA analysis.
    }
    \label{fig:popc_pmf_before_convergence}
\end{figure}

\subsection{Comparison of ABF and Umbrella-Sampling PMFs}

\begin{figure}[H]
    \centering
    \includegraphics[width=0.5\textwidth]{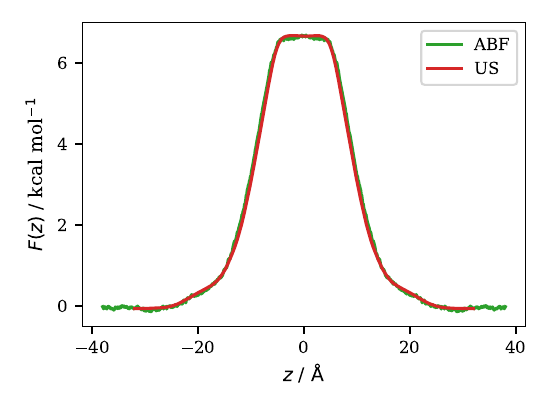}
    \caption{
        PMFs of water permeation through the POPC lipid bilayer obtained from ABF simulations in this work and
        from umbrella sampling in our previous study.\cite{ThomasPrabhakar2025} The two PMF estimates are in close agreement.
        Both systems contain the same number of lipids, but the present POPC/water system has a bulk-water region that is \SI{12}{\angstrom} thicker than in the previous study.\cite{ThomasPrabhakar2025}
    }
    \label{fig:popc_fe_us_abf}
\end{figure}

\subsection{Extended Propagator Comparisons}

\begin{figure}[H]
    \centering
    \includegraphics[width=\textwidth]{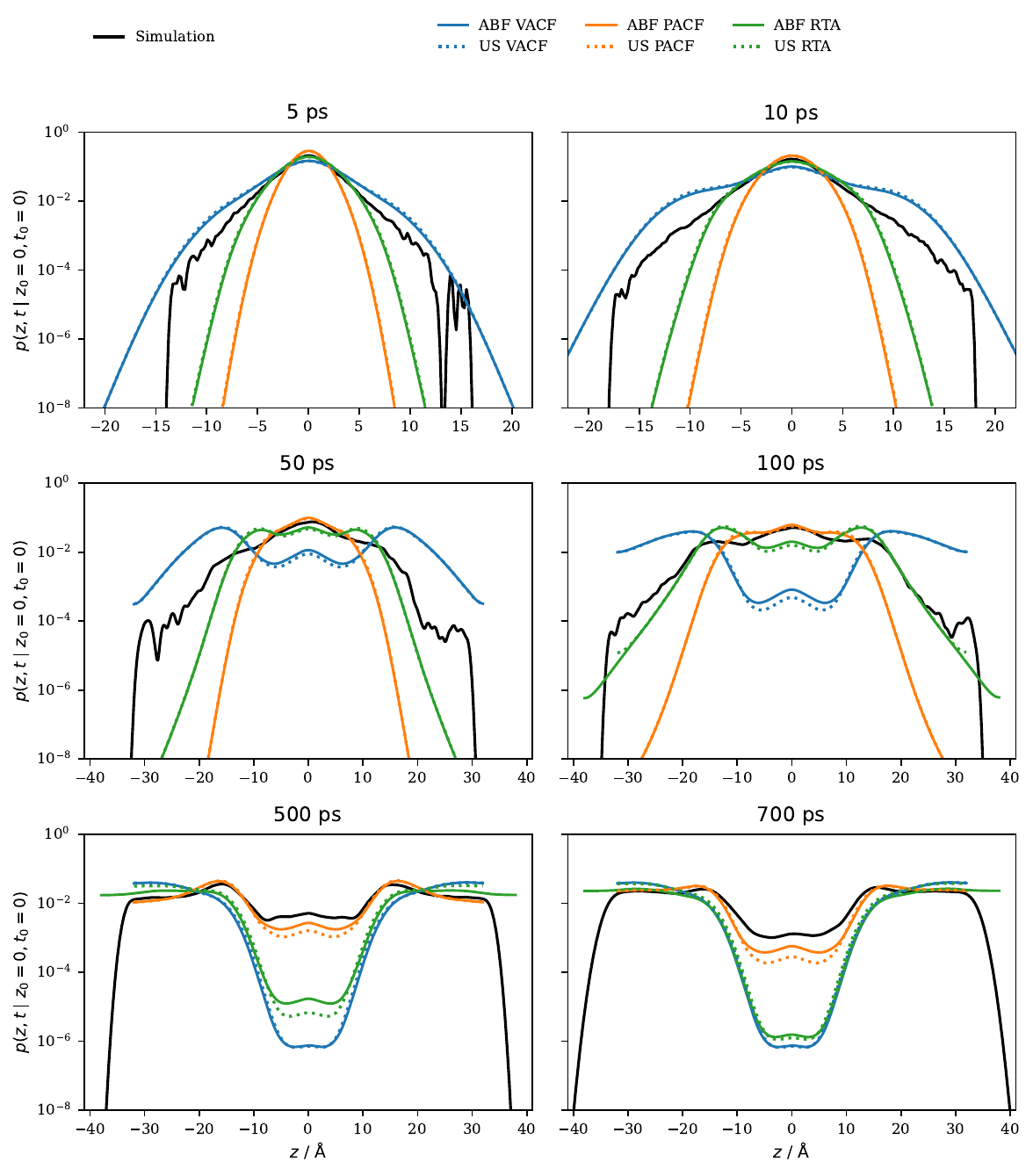}
    \caption{
        POPC/water propagators, \(p(z, t \,|\, z_0, t_0)\), at lag times ranging from
        \(t = \SI{5}{\pico\second}\) to \(t = \SI{700}{\pico\second}\), obtained from simulations and
        diffusion models using the RTA-, PACF-, and VACF-derived diffusivity profiles. For each diffusivity
        profile, predictions obtained using the common ABF-derived estimate and the common US-derived estimate are shown.
    }
    \label{fig:popc_all_propagators}
\end{figure}

To assess sensitivity to the PMF estimate, each RTA-, PACF-, and VACF-derived diffusivity profile was
combined separately with the ABF- and US-derived PMF estimates. Figure~\ref{fig:popc_all_propagators} shows the resulting model
predictions over the complete range of lag times. The relative performance of the three diffusivity profiles is
unchanged between the two PMF estimates: the VACF model provides the closest agreement at very short lag times, the RTA
model performs best over an intermediate range, and the PACF model gives somewhat closer agreement at the longest
lag times. Thus, the relative ranking of the diffusivity profiles is unchanged by the choice of PMF estimate.

\section{Solute Permeation through the Stratum Corneum Membrane}

\subsection{ABF Window Definitions and Sampled Time Intervals}

As shown in our previous study, obtaining well-converged PMFs for the SC system is nontrivial.\cite{ThomasPrabhakar2025}
Accordingly, the CV was divided into \num{7} sequentially overlapping windows with a spacing of \qty{5}{\angstrom},
as summarized in Table~\ref{tab:sc_window}. The initial configurations for each window were extracted from trajectory frames of a
preliminary single-window ABF simulation in which the solute occupied the corresponding target position along the
CV. The solutes were confined within each window using flat-bottomed harmonic walls with a force constant of
\SI{5}{\kcal\per\mol\per\angstrom\squared}. The resulting PMFs from the individual windows were aligned in
their overlap regions using additive constants and then combined into a single continuous PMF.

\begin{table}[H]
    \centering
    \caption{
        Window definitions based on the collective-variable spacing and the corresponding data included for the
        residence-time analysis.
    }
    \label{tab:sc_window}
    \begin{tabular*}{\columnwidth}{@{\extracolsep{\fill}}l
        l
        S[table-format=3.0]
        c
        S[table-format=3.0]}
        \toprule
        Window & {$z$ (\si{\angstrom})} & {Water (\si{\nano\second})} &
        {Acetone (\si{\nano\second})} & {6-MHO (\si{\nano\second})} \\
        \midrule
        1 & -45 to -30 & 320 & 280 & 160 \\
        2 & -35 to -20 & 320 & 280 & 160 \\
        3 & -25 to -5  & 320 & 280 & 280 \\
        4 & -10 to 10  & 280 & 320 & 160 \\
        5 & 5 to 25    & 320 & 280 & 280 \\
        6 & 20 to 35   & 320 & 280 & 160 \\
        7 & 30 to 45   & 320 & 280 & 160 \\
        \bottomrule
    \end{tabular*}
\end{table}

\subsection{Window-Wise PMF Convergence}

\begin{figure}[H]
    \centering
    \includegraphics[width=\textwidth]{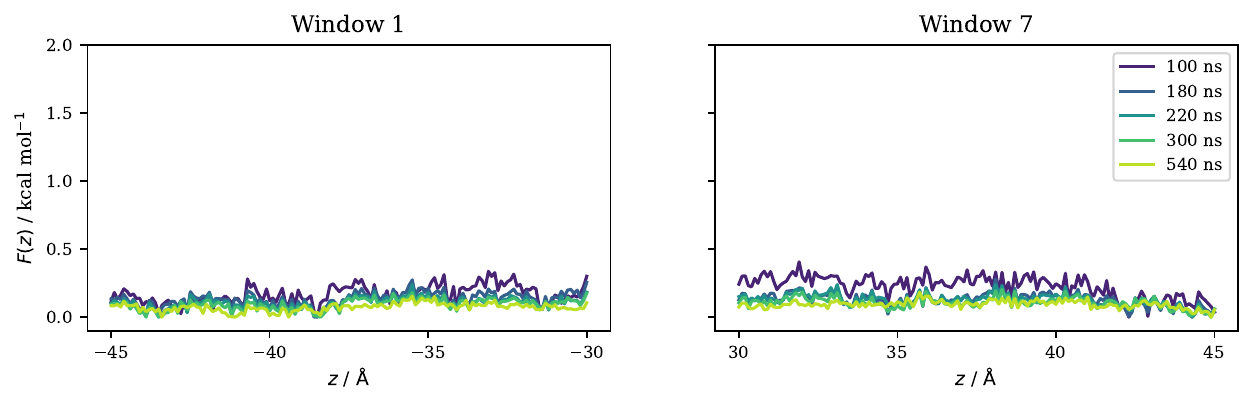}

    \includegraphics[width=\textwidth]{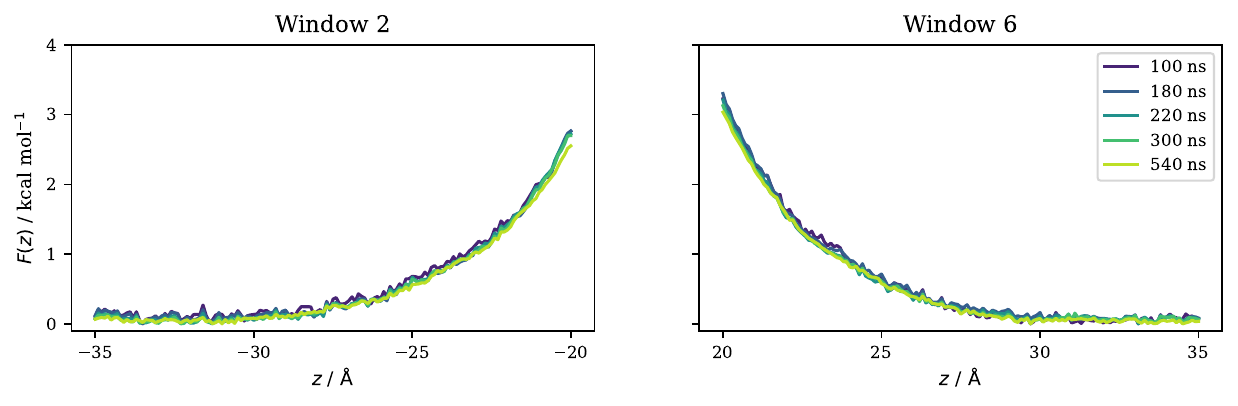}

    \includegraphics[width=\textwidth]{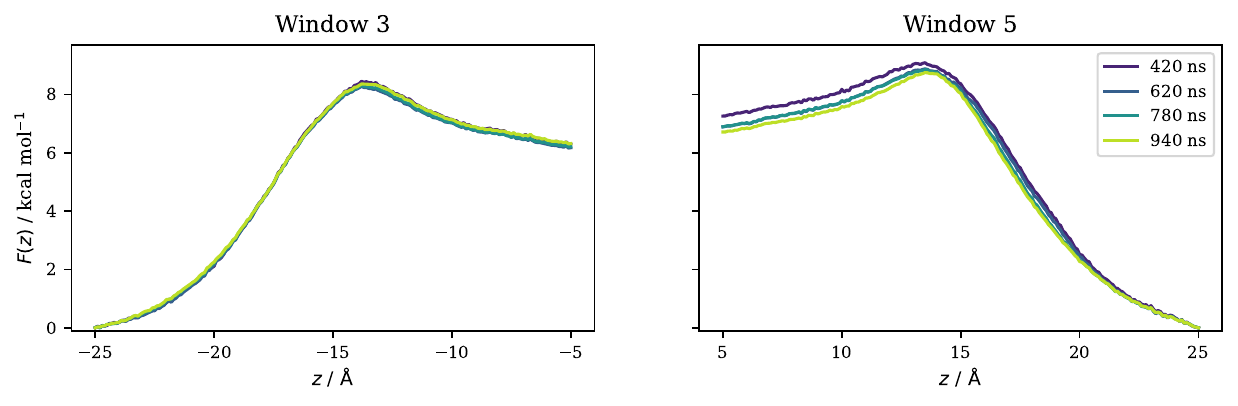}

    \includegraphics[width=0.5\textwidth]{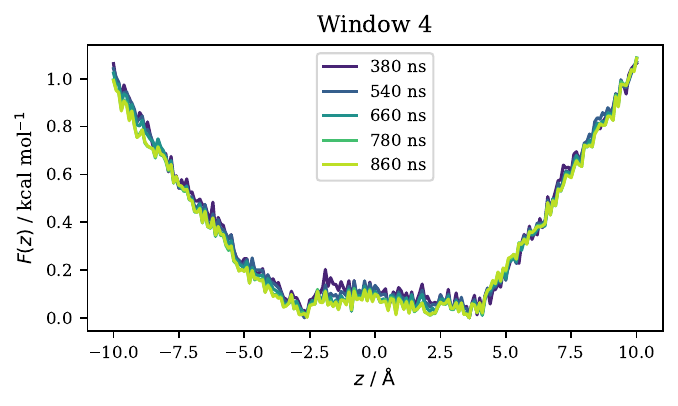}

    \caption{
        Window-wise convergence of the PMFs for water permeation through the SC membrane.
    }
    \label{fig:sc_water_all}
\end{figure}

\begin{figure}[H]
    \centering
    \includegraphics[width=\textwidth]{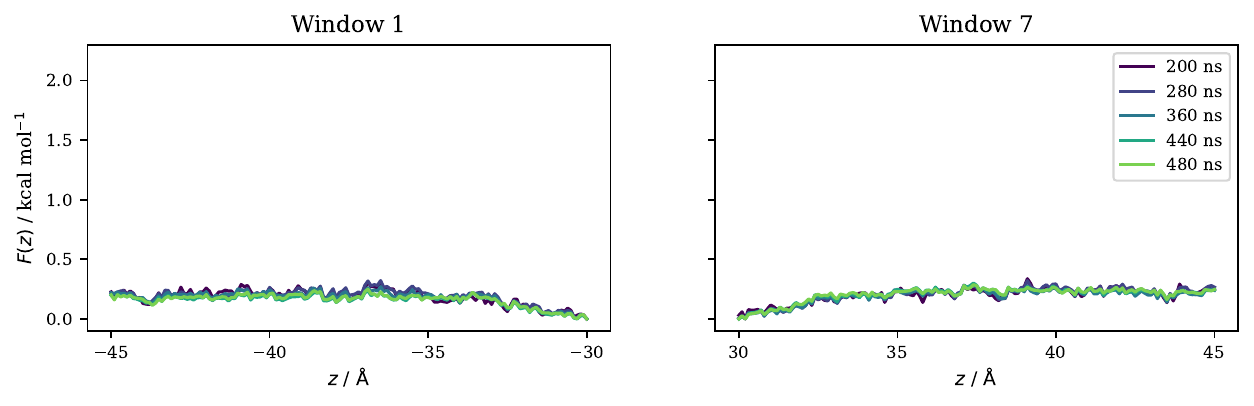}

    \includegraphics[width=\textwidth]{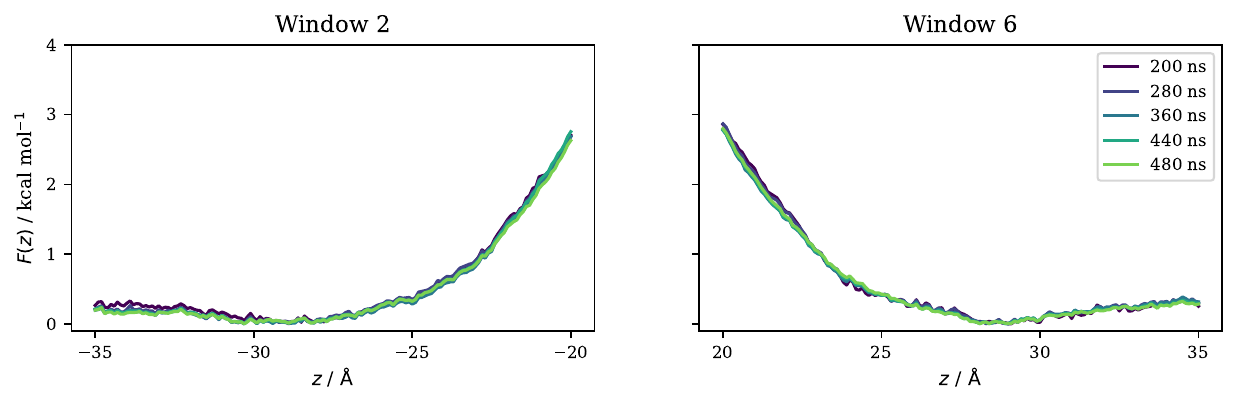}

    \includegraphics[width=\textwidth]{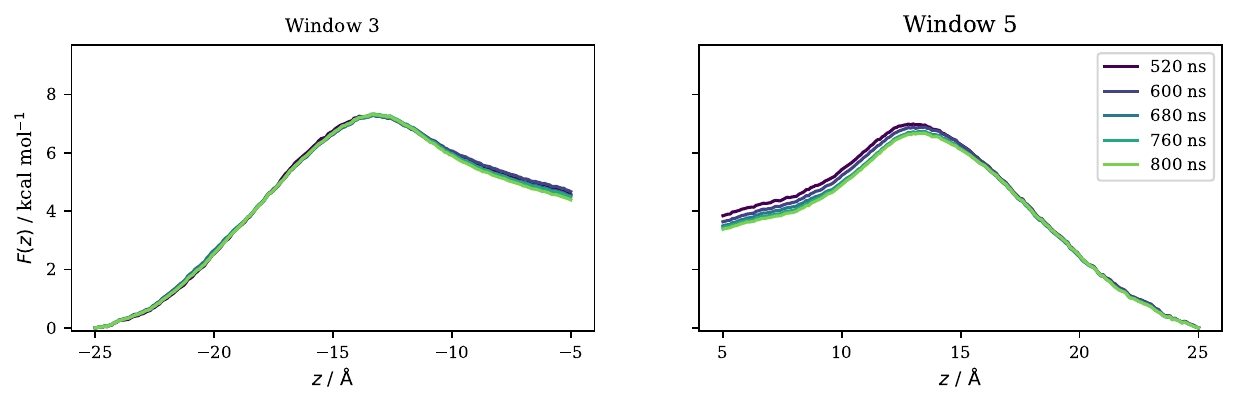}

    \includegraphics[width=0.5\textwidth]{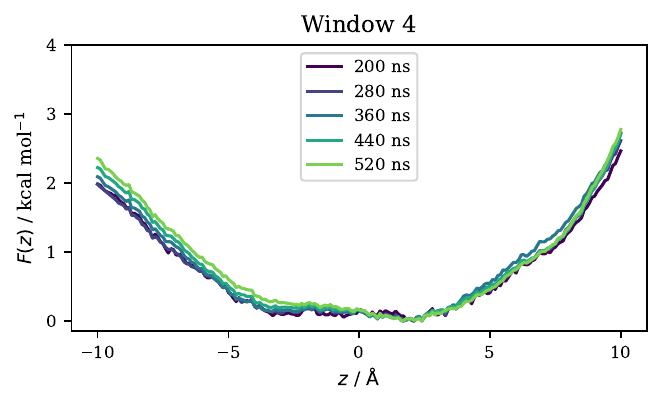}

    \caption{
        Window-wise convergence of the PMFs for acetone permeation through the SC membrane obtained from ABF
        simulations. The RTA analysis was carried out using the simulation data specified in the legends.
    }
    \label{fig:ac_ac_all}
\end{figure}

\begin{figure}[H]
    \centering
    \includegraphics[width=\textwidth]{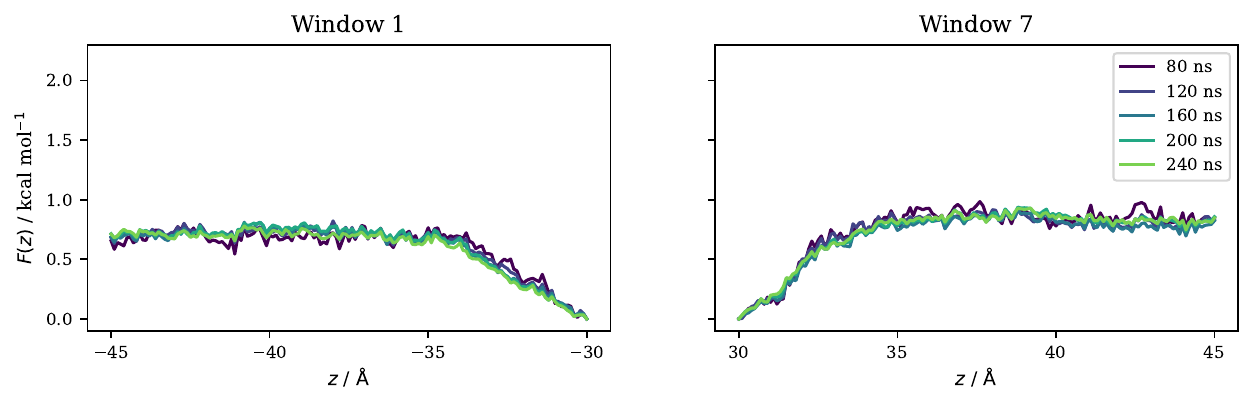}

    \includegraphics[width=\textwidth]{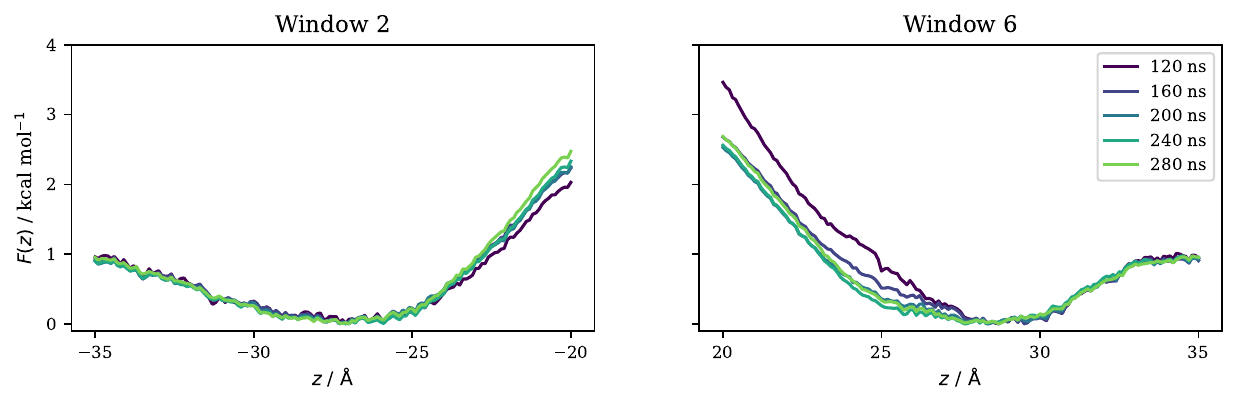}

    \includegraphics[width=\textwidth]{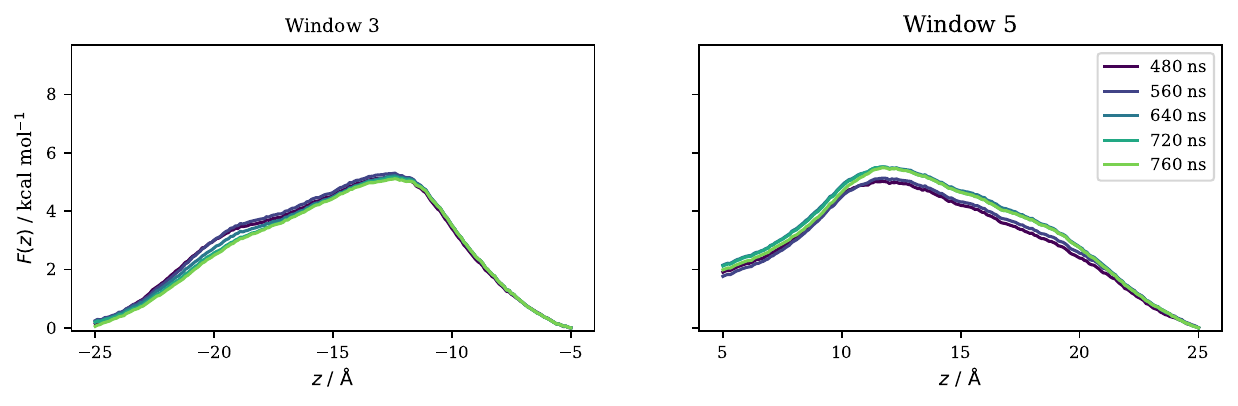}

    \includegraphics[width=0.5\textwidth]{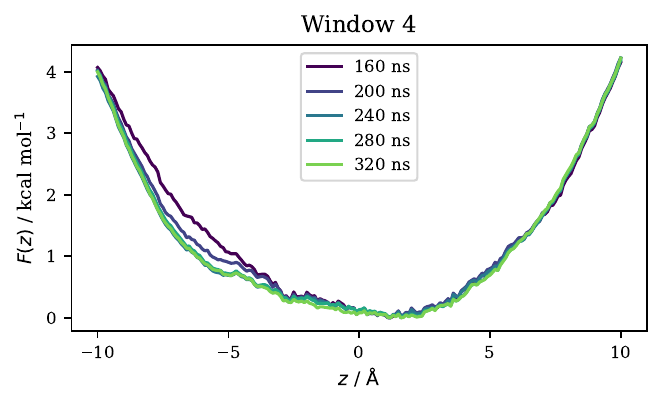}

    \caption{
        Window-wise convergence of the PMFs for 6-MHO permeation through the SC membrane obtained from ABF
        simulations. The RTA analysis was carried out using the simulation data specified in the legends.
    }
    \label{fig:sc_mho_all}
\end{figure}

\subsection{Comparison of PMF Estimates from Different Sampling Methods}

\begin{figure}[H]
    \centering
    \includegraphics[width=0.5\textwidth]{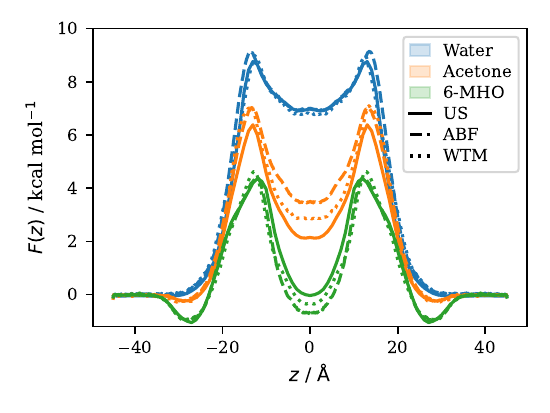}
    \caption{
        PMFs of solute permeation through the SC membrane obtained from ABF simulations in this
        work and from umbrella sampling (US) and well-tempered metadynamics (WTM) simulations in our previous
        study.\cite{ThomasPrabhakar2025} The PMFs are consistent for water. For acetone and 6-MHO, differences of
        less than \SI{1}{\kcal\per\mol} are observed, primarily near the membrane center.
    }
    \label{fig:sc_pmfs}
\end{figure}

\subsection{Extended Propagator Comparisons}

\begin{figure}[H]
    \centering
    \includegraphics[width=\textwidth]{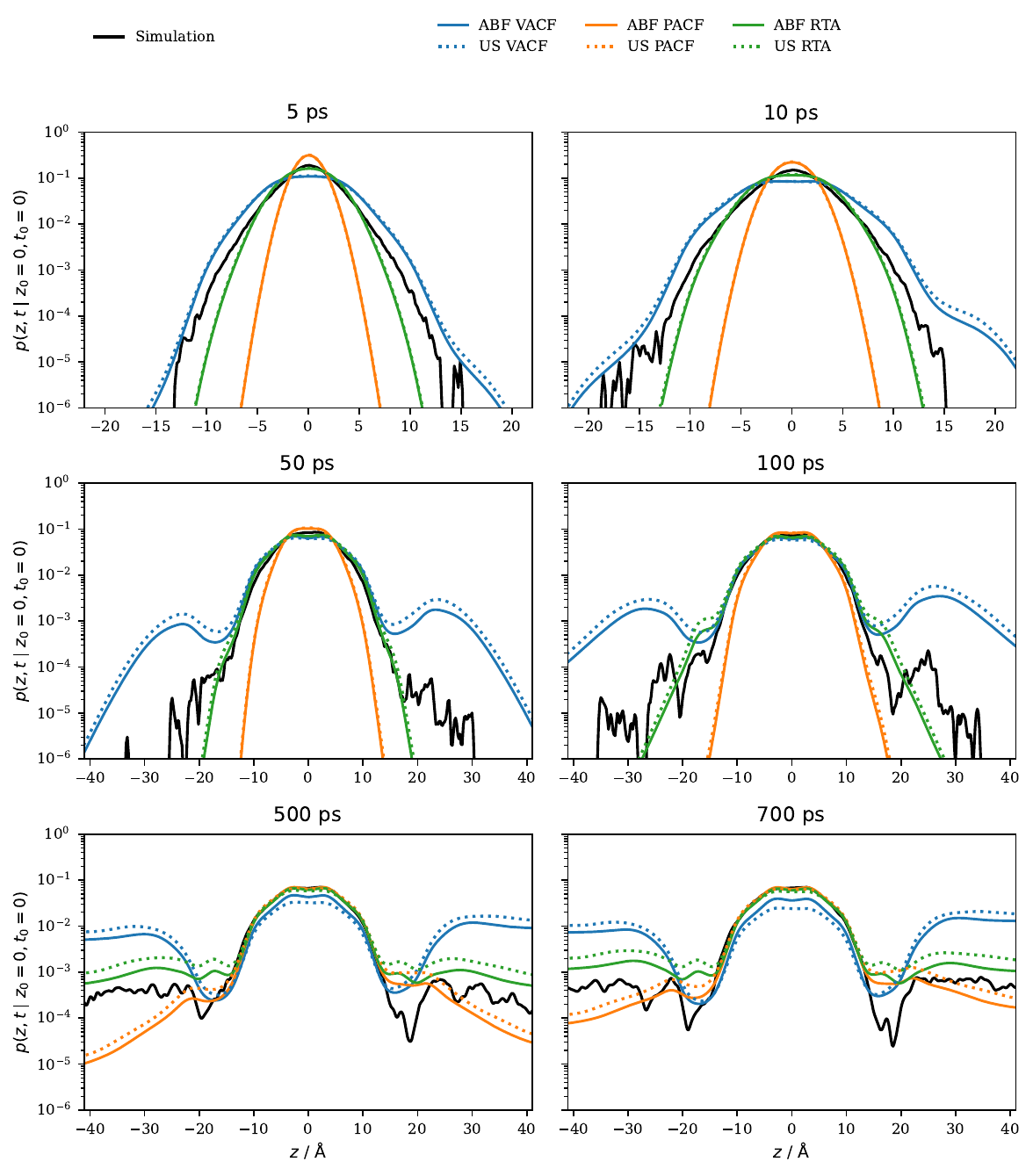}
    \caption{
        SC/water propagators, \(p(z, t \,|\, z_0, t_0)\), at lag times ranging from
        \(t = \SI{5}{\pico\second}\) to \(t = \SI{700}{\pico\second}\), obtained from simulations and diffusion
        models using the RTA-, PACF-, and VACF-derived diffusivity profiles. For each diffusivity profile,
        predictions obtained using the common ABF-derived estimate and the common US-derived estimate are shown.
    }
    \label{fig:sc_water_all_propagators}
\end{figure}

\begin{figure}[H]
    \centering
    \includegraphics[width=\textwidth]{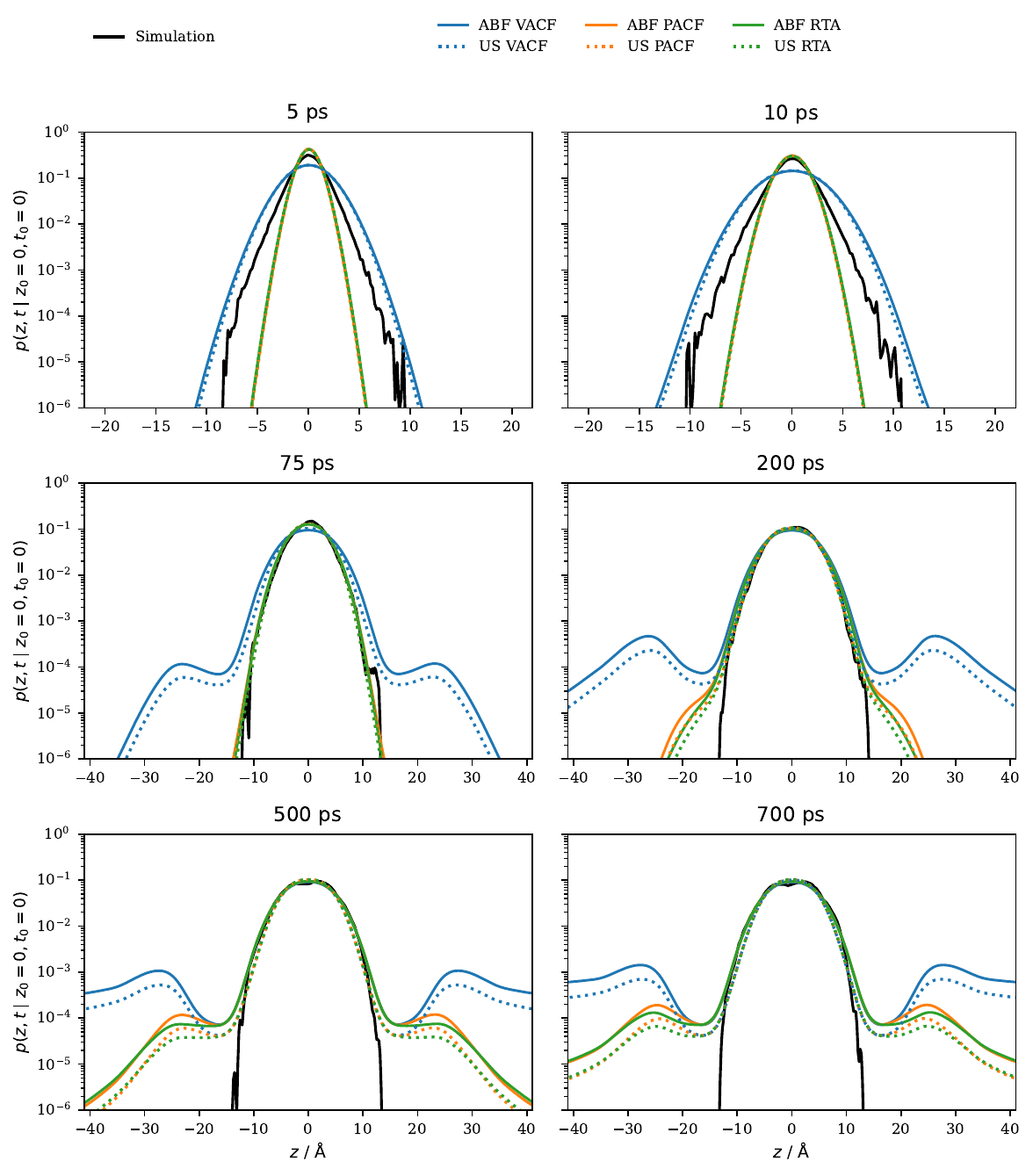}
    \caption{
        SC/acetone propagators, \(p(z, t \,|\, z_0, t_0)\), at lag times ranging from
        \(t = \SI{5}{\pico\second}\) to \(t = \SI{700}{\pico\second}\), obtained from simulations and diffusion
        models using the RTA-, PACF-, and VACF-derived diffusivity profiles. For each diffusivity profile,
        predictions obtained using the common ABF-derived estimate and the common US-derived estimate are shown.
    }
    \label{fig:sc_acetone_all_propagators}
\end{figure}

\begin{figure}[H]
    \centering
    \includegraphics[width=\textwidth]{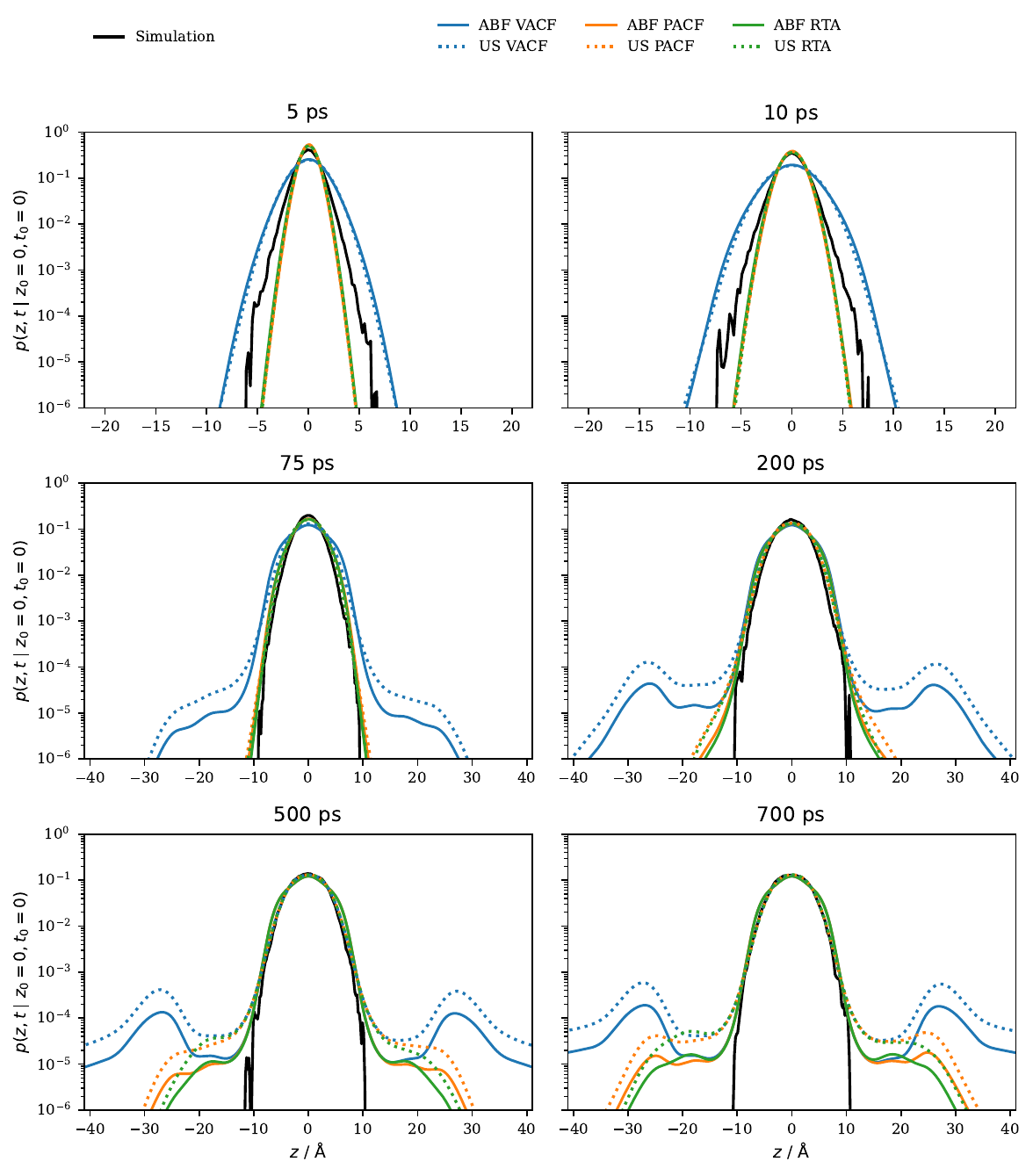}
    \caption{
        SC/6-MHO propagators, \(p(z, t \,|\, z_0, t_0)\), at lag times ranging from
        \(t = \SI{5}{\pico\second}\) to \(t = \SI{700}{\pico\second}\), obtained from simulations and diffusion
        models using the RTA-, PACF-, and VACF-derived diffusivity profiles. For each diffusivity profile,
        predictions obtained using the common ABF-derived estimate and the common US-derived estimate are shown.
    }
    \label{fig:sc_6mho_all_propagators}
\end{figure}

As for POPC, each SC diffusivity profile was combined separately with the ABF- and US-derived PMF estimates. The resulting
propagators are compared with the unbiased MD reference distributions in
Figures~\ref{fig:sc_water_all_propagators}--\ref{fig:sc_6mho_all_propagators}. Substituting the US-derived PMF estimate for the
ABF-derived estimate does not change the relative performance of the diffusivity profiles: the RTA model gives the closest agreement for water, the RTA and
PACF models perform comparably for acetone and 6-MHO, and the VACF models predict systematically broader spreading.

\section{Permeability Analysis}

\subsection{Definition of the Permeability Integration Bounds}

\begin{figure}[H]
    \centering
    \includegraphics[width=0.5\textwidth]{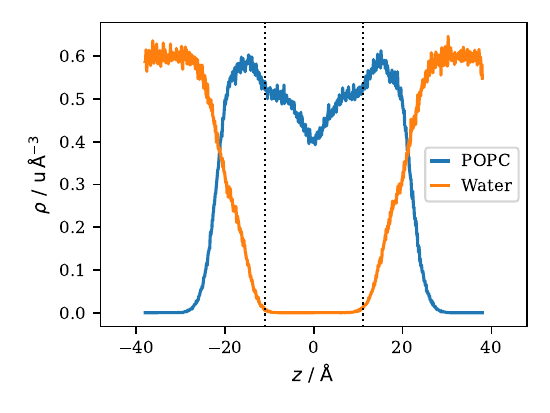}
    \caption{
        Water density profile of the POPC bilayer. The dotted vertical line indicates the cutoff used to define
        the integration bound for the permeability calculation.
    }
    \label{fig:popc_density}
\end{figure}

To estimate permeabilities for the POPC bilayer and SC membrane within the inhomogeneous
solubility--diffusion (ISD) framework, we defined the membrane half-thickness, $h/2$, from the water density
profile as the position beyond which the water density becomes negligibly small. Using this criterion, we chose
$h/2 = \qty{11.0}{\angstrom}$ for POPC and $h/2 = \qty{17.0}{\angstrom}$ for SC. For POPC, this definition was
applied analogously to the procedure used previously for SC.\cite{ThomasPrabhakar2026}
Figure~\ref{fig:popc_density} shows the water density profile for the POPC system together with the resulting
integration cutoff.

\subsection{Supplementary Permeability Coefficients}

Table~\ref{tab:permeability} reports permeabilities obtained from the full factorial combination of the two PMF estimates
and three diffusivity profiles. Comparisons across a row hold the PMF fixed and therefore show the influence of the
diffusivity profile. Comparisons between the ABF and US rows for a fixed diffusivity quantify sensitivity to the
PMF estimate. The ordering associated with the three diffusivity profiles is unchanged between the two estimates for every
system, although the absolute permeability coefficients, particularly for the SC systems, exhibit appreciable PMF
sensitivity.

\begin{table}[H]
    \centering
    \caption{
        Membrane permeabilities, reported in \si{\centi\meter\per\second}, obtained using the
        inhomogeneous solubility--diffusion model. Each RTA-, PACF-, and VACF-derived diffusivity profile was
        combined separately with the ABF- and US-derived PMF estimates.
    }
    \label{tab:permeability}
    \begin{threeparttable}
    \begin{tabular*}{\columnwidth}{@{\extracolsep{\fill}}llcccc}
        \toprule
        System & PMF & RTA & PACF & VACF & Literature \\
        \midrule
        POPC/water\tnote{a} & ABF & \num{6.03e-3} & \num{2.51e-3} & \num{1.39e-2} & \num{1.40e-3} \\
        POPC/water & US & \num{6.20e-3} & \num{2.34e-3} & \num{1.30e-2} & --- \\
        \addlinespace
        SC/water\tnote{b} & ABF & \num{3.81e-5} & \num{3.28e-5} & \num{2.53e-4} & \num{1.88e-5} \\
        SC/water & US & \num{8.19e-5} & \num{5.65e-5} & \num{4.15e-4} & --- \\
        \addlinespace
        SC/acetone & ABF & \num{6.73e-4} & \num{8.00e-4} & \num{5.37e-3} & --- \\
        SC/acetone & US & \num{2.21e-3} & \num{2.66e-3} & \num{1.79e-2} & --- \\
        \addlinespace
        SC/6-MHO & ABF & \num{3.94e-2} & \num{5.02e-2} & \num{3.11e-1} & --- \\
        SC/6-MHO & US & \num{3.60e-2} & \num{4.55e-2} & \num{2.82e-1} & --- \\
        \bottomrule
    \end{tabular*}
    \begin{tablenotes}
        \item[a] Literature value from Venable \textit{et al.}\cite{venable_molecular_2019} Their simulations were performed at \SI{303.15}{\kelvin}.
        \item[b] Literature value from Reuter \textit{et al.}\cite{reuter_presence_2025} For a broader comparison across force fields and methods, see Thomas \textit{et al.}\cite{ThomasPrabhakar2026}
    \end{tablenotes}
    \end{threeparttable}
\end{table}

\section{Methodological Validation of the Residence-Time Approach}

\subsection{Local Dynamical Statistics}

To further examine the assumptions underlying the residence-time analysis (RTA),
we consider local dynamical statistics computed from trajectory segments confined
to spatial intervals of width $L$. Specifically, we compare the local
mean-square displacement (MSD) and the survival probability $S(t)$ for
representative bulk and membrane-center regions in the SC/water system.

\subsubsection{Local Mean-Square Displacements}

The local MSD was computed from contiguous trajectory segments that remain
within the interval throughout the observation time. Because the dynamics are
confined to a finite domain, the MSD cannot grow indefinitely and necessarily
deviates from linear behavior on timescales comparable to the mean residence
time $\tau_r$. Consequently, deviations from linearity on these timescales
cannot, by themselves, be interpreted as evidence against an effective
Smoluchowski description.

\begin{figure}[H]
    \centering
    \includegraphics[width=\textwidth]{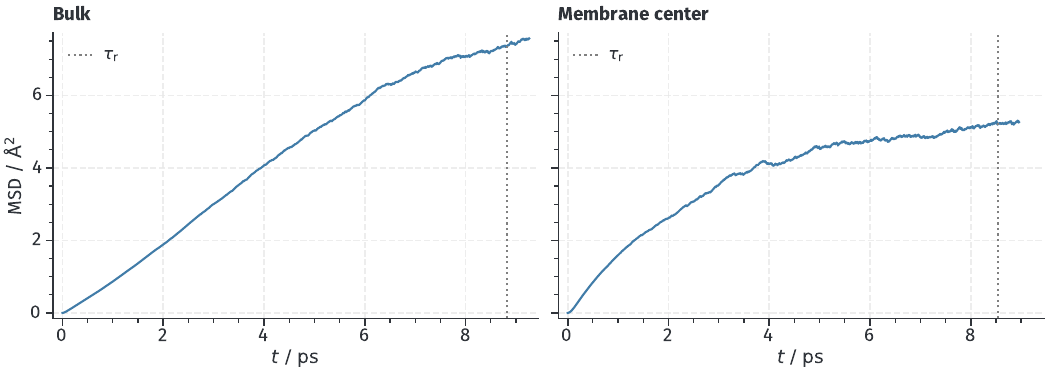}
    \caption{
    Local mean-square displacements computed from trajectory segments that
    remain within representative bulk and membrane-center intervals of width
    $L$ in the SC/water system. The vertical dashed lines indicate the
    corresponding mean residence times $\tau_r$.
    }
    \label{fig:local_msd}
\end{figure}

The resulting MSDs are shown in Figure~\ref{fig:local_msd}. Both the bulk and
membrane-center regions exhibit an approximately linear regime at intermediate
times, followed by the expected saturation as the particle explores the entire
interval. The membrane-center region displays more pronounced deviations from
linearity than the bulk. However, because the MSD is strongly influenced by the
finite extent of the observation interval, it provides only an indirect test of
the assumptions underlying the RTA and does not determine whether the
corresponding first-passage process is adequately described by an effective
Markovian diffusion model.

\subsubsection{Survival Probabilities}

A more direct test of the assumptions underlying the RTA is provided by the
survival probability,

\begin{equation}
S(t)=P(T>t),
\end{equation}

where $T$ denotes the remaining time until first exit from the interval.

For diffusion in a finite interval with absorbing boundaries, the survival
probability is given by a sum of exponentially decaying eigenmodes,

\begin{equation}
S(t)=\sum_n a_n e^{-\lambda_n t},
\end{equation}

where $\lambda_n$ are the eigenvalues of the corresponding Smoluchowski
operator.\cite{Redner2001} At sufficiently long times, the lowest eigenmode
dominates, so that the survival probability approaches a mono-exponential
decay.

For the present comparison, we evaluate the analytical survival probability for
homogeneous one-dimensional Smoluchowski diffusion with constant diffusivity.
The diffusivity is fixed by the RTA relation,

\begin{equation}
D=\frac{L^2}{12\tau_r},
\end{equation}

such that the comparison contains no adjustable parameters.

\begin{figure}[H]
    \centering
    \includegraphics[width=\textwidth]{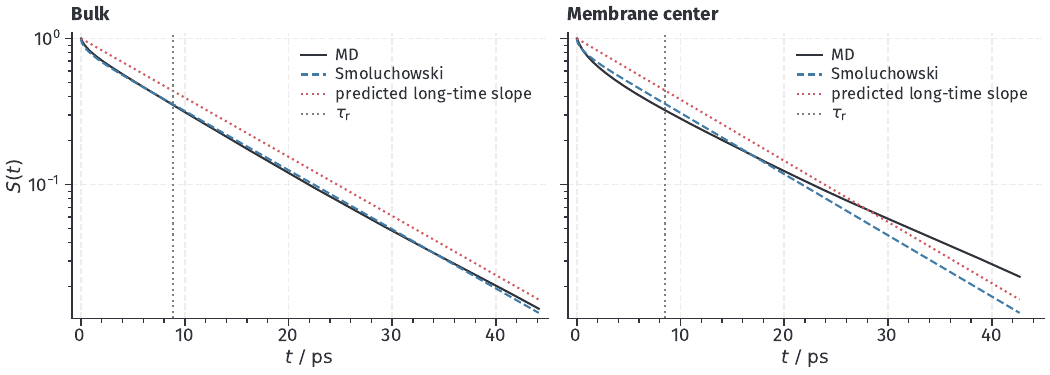}
    \caption{
    Survival probabilities obtained from MD for representative bulk and
    membrane-center intervals in the SC/water system compared with the
    analytical prediction for homogeneous one-dimensional Smoluchowski
    diffusion using the RTA diffusivity,
    $D=L^2/(12\tau_r)$, without additional fitting.
    }
    \label{fig:local_survival}
\end{figure}

Figure~\ref{fig:local_survival} compares the survival probabilities obtained
directly from MD with the corresponding Smoluchowski predictions. In the bulk
region, close agreement is observed over the entire time range, indicating
that the first-passage statistics are well described by an effective diffusive
process.

In contrast, the membrane-center region exhibits systematic deviations from the
homogeneous Smoluchowski prediction. The MD survival probability decays more
rapidly at short times but more slowly at long times, indicating a broader
distribution of residence times than predicted by homogeneous diffusion while
preserving the same mean residence time. Since the analytical prediction is
constructed from the RTA diffusivity, both distributions necessarily yield the
same average residence time, but differ in the detailed distribution of
first-passage events.

The approximately mono-exponential long-time decay observed for the MD
survival probability is consistent with the dominance of a single slow
relaxation mode expected for diffusion in a finite interval. However, such
behavior is not unique to Markovian dynamics and may also arise from spatially
heterogeneous diffusivity or free-energy variations within the interval.
Consequently, the observed deviations from the homogeneous Smoluchowski
prediction cannot be uniquely attributed to memory effects.

Overall, these results suggest that the RTA diffusivity should be interpreted
as an effective diffusivity characterizing transport over the length scale
defined by the residence interval, rather than as a unique local transport
coefficient governing all dynamical observables.

Finally, we emphasize that both the local MSD and the survival probability
probe only selected aspects of the projected dynamics. The most stringent
validation of the reduced Smoluchowski description is provided by comparison of
the full propagator with unbiased MD simulations, as presented in the main
text. Because the propagator determines the complete time-dependent transition
probability, it provides a more global characterization of diffusive transport
than local dynamical statistics.

\subsubsection{Additional systems}

\begin{figure}[H]
    \centering
    \includegraphics[width=\textwidth]{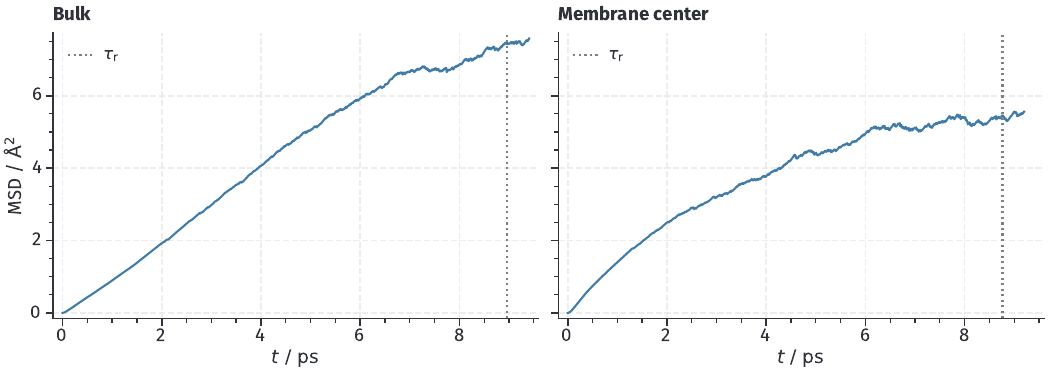}
    \vspace{0.5em}
    \includegraphics[width=\textwidth]{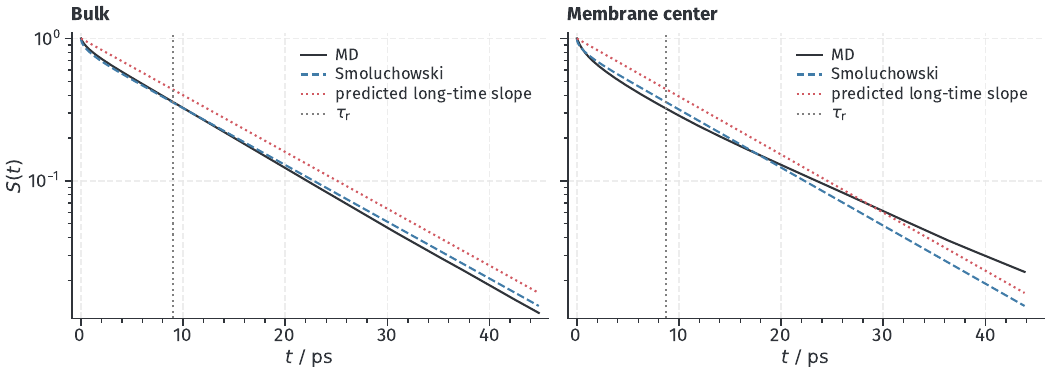}
    \caption{
    Local dynamical statistics for water permeation across the POPC membrane.
    Top: mean-square displacements computed from trajectory segments confined
    to representative bulk and membrane-center intervals. Bottom:
    corresponding survival probabilities obtained from MD compared with the
    analytical prediction for homogeneous Smoluchowski diffusion using the RTA
    diffusivity, $D=L^2/(12\tau_r)$.
    }
    \label{fig:popc_local_statistics}
\end{figure}

\begin{figure}[H]
    \centering
    \includegraphics[width=\textwidth]{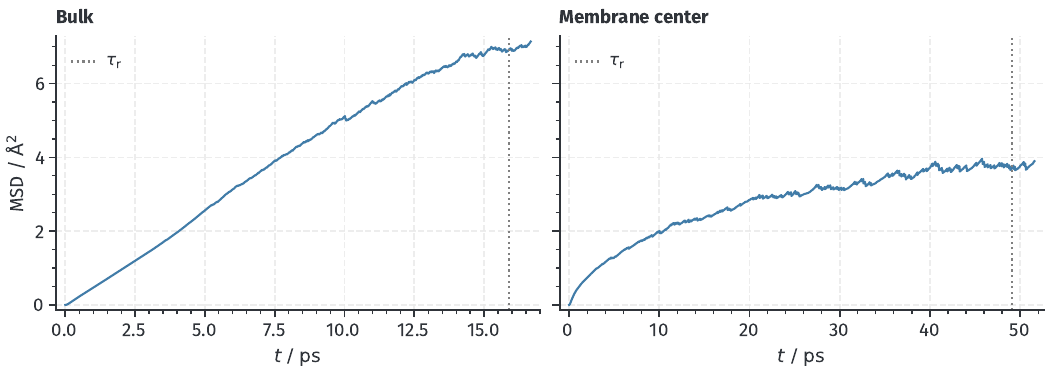}
    \vspace{0.5em}
    \includegraphics[width=\textwidth]{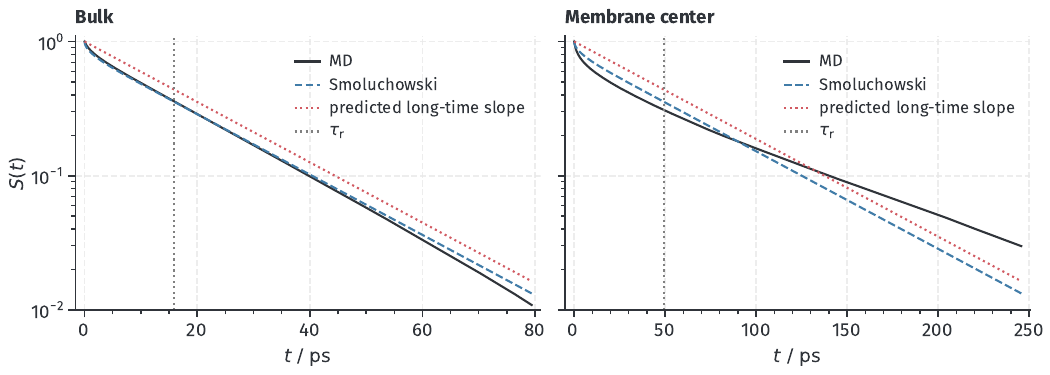}
    \caption{
    Same as Figure~\ref{fig:popc_local_statistics}, but for acetone
    permeation across the SC membrane.
    }
    \label{fig:sc_acetone_local_statistics}
\end{figure}

\begin{figure}[H]
    \centering
    \includegraphics[width=\textwidth]{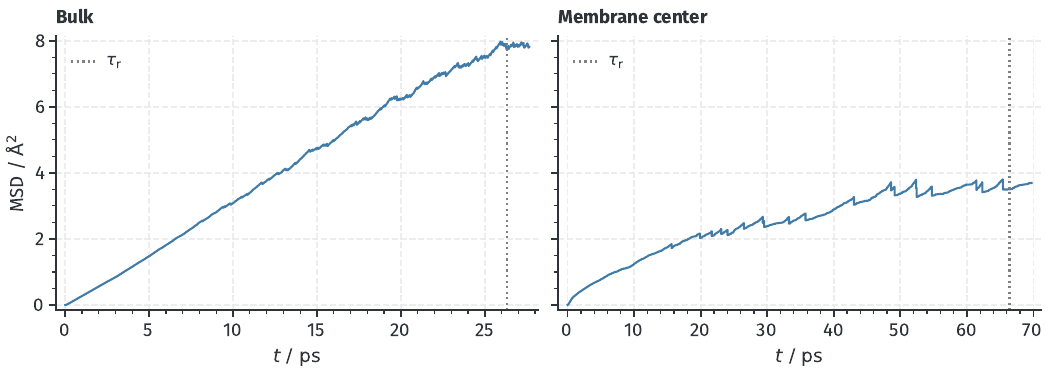}
    \vspace{0.5em}
    \includegraphics[width=\textwidth]{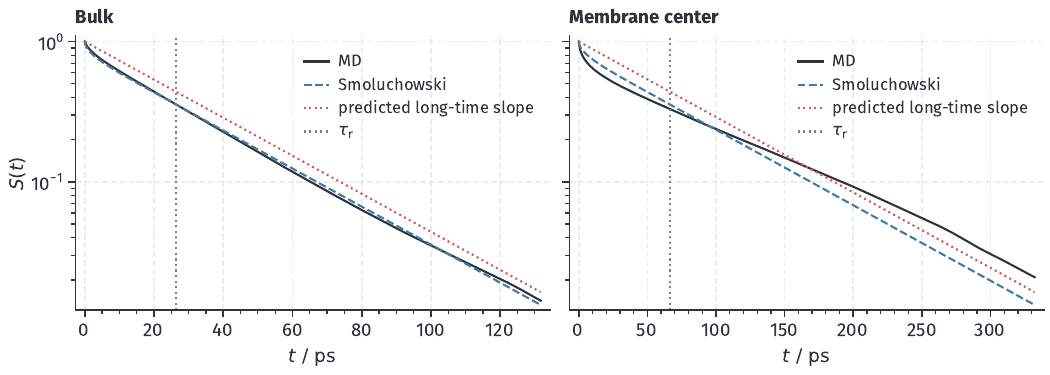}
    \caption{
    Same as Figure~\ref{fig:popc_local_statistics}, but for 6-MHO
    permeation across the SC membrane.
    }
    \label{fig:sc_6mho_local_statistics}
\end{figure}

Equivalent analyses for the remaining membrane systems considered in this work
are shown in Figures~\ref{fig:popc_local_statistics},
\ref{fig:sc_acetone_local_statistics}, and
\ref{fig:sc_6mho_local_statistics}. In all cases, the local MSD exhibits the
expected saturation on timescales comparable to the mean residence time,
confirming that deviations from linearity are a generic consequence of
confinement within the residence interval. Likewise, the survival probability
is generally well described by the homogeneous Smoluchowski prediction in
bulk-like regions, while systematic deviations are observed within the membrane
interior. Despite these deviations, the RTA correctly reproduces the mean
residence time by construction and should therefore be interpreted as providing
an effective diffusivity on the scale of the residence interval. Since the full
propagator comparison presented in the main text constitutes a strictly
stronger validation of the reduced Smoluchowski model, we do not analyze these
local statistics further.

\subsection{Uncertainty Quantification for Correlated Residence-Time Data}

Reliable estimation of statistical uncertainties from correlated simulation data is a long-standing challenge. In the main text, uncertainties in the mean residence times were estimated using the automated blocking procedure of Jonsson,\cite{Jonsson2018} which builds upon the blocking transformation of Flyvbjerg and Petersen.\cite{FlyvbjergPetersen1989} This approach estimates the asymptotic variance directly from the data without requiring an explicit model for the temporal correlation structure.

Recently, McCluskey \emph{et al.}\cite{McCluskey2025AccurateEstimation} proposed an efficient framework for estimating bulk diffusion coefficients and their statistical uncertainties from a single simulation trajectory. In addition to their main estimator, they discuss a variance-normalization strategy in which the variance of the mean is obtained by dividing the variance of the underlying observable by the effective number of statistically independent samples. For mean-squared displacement (MSD) analyses, this procedure is particularly effective because the correlation structure arising from overlapping displacement windows is known analytically, allowing the effective number of independent observations to be estimated directly.

Residence-time data differ fundamentally from MSD data in this respect. Although successive residence times extracted from a trajectory are correlated, the corresponding correlation structure is generally not known \emph{a priori}. Correlations arise both from the use of overlapping starting points within a residence event and from slow dynamical processes in the underlying molecular system, such as persistent membrane fluctuations and hidden degrees of freedom. Consequently, a direct variance-normalization procedure analogous to that proposed for MSD data is not readily available.

To assess the robustness of the uncertainty estimates employed in the present work, we therefore compared the automated blocking procedure with two independent analyses based on the same residence-time data. First, variance estimates were computed after successive Flyvbjerg--Petersen blocking transformations without applying the automated stopping criterion. In each blocking transformation, neighboring observations are averaged pairwise, thereby increasing the effective block size while preserving the sample mean. The variance of the mean residence time is estimated from the variance of the blocked data divided by the remaining number of observations. As short-range correlations are progressively removed, the estimated variance increases toward an asymptotic plateau corresponding to effectively independent blocks.

Second, the residence-time series was partitioned into increasing numbers of non-overlapping contiguous batches. The variance of the mean residence time was estimated from the variance of the corresponding batch means divided by the number of batches. Increasing the number of batches reduces the batch size and therefore moves the estimator toward the result obtained by treating individual residence times as independent. Consequently, the estimated variance decreases with increasing batching level and approaches the asymptotic value only while the batches remain sufficiently large to average over the dominant temporal correlations.

\begin{figure}[H]
    \centering
    \includegraphics[width=\textwidth]{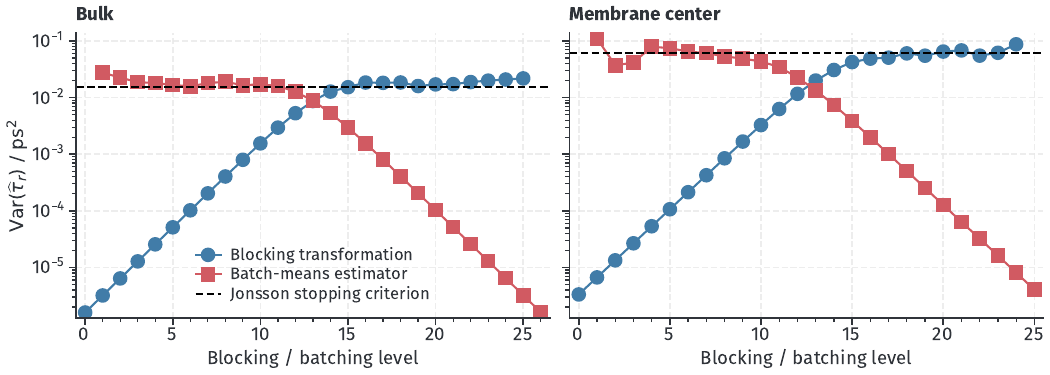}
    \caption{Comparison of variance estimates for the mean residence time obtained from successive Flyvbjerg--Petersen blocking transformations (circles), a non-overlapping batch-means estimator (squares), and the automated blocking procedure of Jonsson (horizontal dashed line). Results are shown for a representative bulk interval (left) and for the membrane center (right). The blocking estimate increases with successive blocking transformations as short-range correlations are removed, whereas the batch-means estimate decreases as the batch size is reduced. Both approaches converge to closely similar asymptotic values, in agreement with the automated blocking estimate.}
    \label{fig:variance}
\end{figure}

The resulting variance estimates for a representative bulk interval and for the membrane center are shown in Figure~\ref{fig:variance}. In both cases, the blocking transformation approaches the asymptotic variance from below, whereas the batch-means estimator approaches the same limiting value from above before decreasing toward the independent-sample limit for very small batch sizes. The automated Jonsson estimate lies within the plateau defined by these two independent analyses and is in excellent agreement with their common asymptotic value.

The close agreement between these conceptually different approaches indicates that the uncertainty estimates reported in the present work are not sensitive to the particular method used to account for temporal correlations. Although the variance-normalization framework of McCluskey \emph{et al.} cannot be transferred directly to residence-time statistics because of the unknown correlation structure, the comparison presented here provides additional evidence that the automated blocking procedure yields reliable estimates of the asymptotic variance for the residence-time data considered in this work.

\subsection{Dependence on the Residence Interval Width}

The residence interval width, $L$, defines the spatial coarse-graining scale of
the residence-time approach and therefore influences the effective diffusivity
extracted from the first-passage statistics. If $L$ is chosen too small, the
residence times become increasingly sensitive to short-time non-diffusive
dynamics and statistical noise. Conversely, if $L$ is too large, spatial
variations of the PMF and diffusivity within the interval may violate the
assumption of locally constant transport properties underlying the RTA.

\begin{figure}[H]
    \centering
    \includegraphics[width=0.5\textwidth]{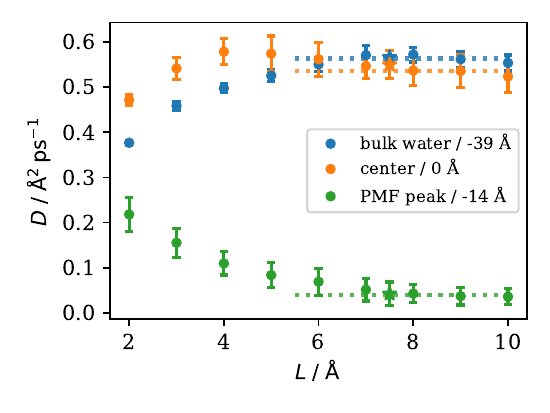}
    \caption{
        Dependence of the RTA diffusivity on the residence interval width $L$
        for three representative regions of the SC/water system: bulk water,
        the membrane center, and the PMF maximum. The plateau observed at
        larger interval widths indicates that the estimated diffusivities are
        only weakly sensitive to the precise choice of $L$ within this
        coarse-graining regime.
    }
    \label{fig:bin_size}
\end{figure}

To examine the dependence of the estimated diffusivity on $L$, we performed the
residence-time analysis using interval widths ranging from
\qty{2}{\angstrom} to \qty{10}{\angstrom} for three representative regions of
the SC/water system: bulk water ($z=\qty{-39}{\angstrom}$), the membrane center
($z=\qty{0}{\angstrom}$), and the PMF maximum
($z=\qty{-14}{\angstrom}$). The resulting diffusivities are shown in
Figure~\ref{fig:bin_size}.

For all three regions, the estimated diffusivity exhibits a plateau within
statistical uncertainty over the interval-width range from
\qty{7.5}{\angstrom} to \qty{10}{\angstrom}. This behavior indicates that the
estimated diffusivity depends only weakly on the precise choice of $L$ within
this coarse-graining regime. The interval width adopted throughout the present
work, $L=\qty{7.5}{\angstrom}$, lies within this plateau and therefore provides
a robust estimate of the effective diffusivity. This conclusion is further
supported by the propagator analysis presented in the main text, which
demonstrates that diffusivities obtained with this choice of $L$ yield a
reduced Smoluchowski description over the relevant range of lag
times.

\bibliography{si}